\documentclass[12pt]{iopart}
\usepackage{amssymb}
\usepackage{iopams}

\usepackage{setstack}

\begin{document}

\title{Geometrical approach to the discrete Wigner function}

\author{A. B. Klimov,  C. Mu\~{n}oz and J.L. Romero}

\address{\dag\ Departamento de F\'{\i }sica, Universidad de
Guadalajara, Revoluci\'{o}n 1500, 44410, Guadalajara, Jal.,
M\'{e}xico.}

\begin{abstract}
We analyze the Wigner function constructed on the basis of the
discrete rotation and displacement operators labelled with
elements of the underlying finite field. We separately discuss the
case of odd and even characteristics and analyze the algebraic
origin of the non uniqueness of the representation of the Wigner
function. Explicit expressions for the Wigner kernel are given in
both cases.
\end{abstract}

\section{Introduction}

The Wootters' construction of the Wigner function \cite{Wootters, Wootters04}
in the discrete phase space for quantum systems whose dimension $d=p^{n}$ is
a power of a prime number has been applied for the analysis of several
problems, mainly related to quantum information processes \cite{Buzek,
Paz1,Paz2,Paz3}. In this case, the phase space is a $d\times d$ grid with\ a
well defined geometrical structure (the so-called finite geometry) \cite{FF}%
, which is a direct consequence of the underlying finite field (Galois
field) structure, so that the phase-space coordinates can be chosen as
elements of the corresponding finite field \cite{Wootters, Wootters04}. The
Wigner function (a discrete symbol of the density matrix) is constructed
using the so-called phase point operators and consists in a highly
non-univoque procedure of association of states in the Hilbert space with
some geometrical structures (lines) in the discrete phase space, is a
substantial disadvantage.

Slightly different, operational approaches to the Wigner function
construction, based on the algebraic structure of the Galois
fields, have been recently \cite{Vourdas05,Rubin05, Klimov05b,
Durt05} proposed for the quantum systems of prime power
dimensions, although, for the prime dimensions such constructions
were also discussed earlier \cite{WFold,Vourdas}.

In the present article we analyze the Wigner function constructed
on the basis of the discrete rotation and displacement operators
labelled with the elements of the underlying finite field. We
separately discuss the case of odd and even characteristics and
analyze the algebraic origin of the non-uniqueness of the
representation of the Wigner function in both cases. The main
difference with the Wootters' method consists in labelling the
elements of the Hilbert space and the operators acting on this
space directly by elements of the finite field \cite{Klimov05a},
so that the possible factorizations of the corresponding operators
appear as a result of choosing different basis in the field, and
are\ not really necessary for the phase space construction.

The paper is organized as follows. In Sec.2 we outline the structure of the
generalized Pauli group and introduce the Wigner mapping as that which
satisfies the Stratonovich-Weyl postulates and shows the phase problem for
the displacement operator. In Sec.3 we analyze the rotation operators for
fields of odd and even characteristics and discuss the discrete phase space
construction. In Sec.4 we briefly outline the reconstruction \ procedure. In
Sec.5 we discuss different possibilities for ordering points in the phase
space with elements of the finite field. In Sec.6 we analyze the relation
between abstract states and states of physical systems. Several useful
relations are proved in Appendices.

\section{General definitions}

In the case of prime power dimensions, $d=p^{n}$, the phase-space
representation can be constructed in a similar way as in the prime
dimensions \cite{Wootters,Vourdas}. We perform the Wigner mapping
in a slightly different manner than in \cite{Wootters04}, by using
the generalized position and momentum operators introduced in
\cite{Klimov05a} (see also \cite{Vourdas05}). In the case of
composite dimensions instead of natural numbers we have to use
elements of the finite field $GF(d)$ to label states of the system
and operators acting on the corresponding Hilbert space. In
particular, we will denote as $\left\vert \alpha \right\rangle ,\
\alpha \in GF\left( d\right) $ an orthonormal basis in the Hilbert
space of the quantum system, $\langle \alpha \left\vert \beta
\right\rangle =\delta _{\alpha ,\beta }$. Operationally, the
elements of the basis can be labelled by powers of primitive
elements (see Appendix A). These vectors will be considered as
eigenvectors of the generalized position operator which belong to
the generalized Pauli group. The generators of this group, usually
called generalized position and momentum operators, are defined as
follows
\begin{eqnarray}
Z_{\beta }\left\vert \alpha \right\rangle &=&\chi \left( \alpha \beta
\right) \left\vert \alpha \right\rangle ,\quad X_{\beta }\left\vert \alpha
\right\rangle =\left\vert \alpha +\beta \right\rangle ,\quad \alpha ,\beta
\in GF\left( d\right) ,  \label{XZgf} \\
Z_{\beta }^{\dagger } &=&Z_{-\beta },\quad X_{\beta }^{\dagger
}=X_{-\beta },
\end{eqnarray}%
so that%
\begin{equation*}
Z_{\alpha }X_{\beta }=\chi \left( \alpha \beta \right) X_{\beta }Z_{\alpha },
\end{equation*}%
where $\chi (\theta )$ is an additive character \cite{FF}%
\begin{equation}
\chi (\theta )=\exp \left[ \frac{2\pi i}{p}\mathrm{tr}\left( \theta \right) %
\right] ,  \label{ch}
\end{equation}%
and the trace operation, which maps elements of $GF\left( d\right) $ into
the prime field $GF\left( p\right) \simeq \mathcal{Z}_{p}$ is defined as%
\begin{equation*}
\mathrm{tr}\left( \theta \right) =\theta +\theta ^{p}+\theta
^{p^{2}}+...+\theta ^{p^{n-1}},
\end{equation*}%
this operation leaves the elements of the prime field (see Appendix A)
invariant. The characters (\ref{ch}) satisfy the following important
properties:

\begin{equation}
\sum_{\alpha \in GF\left( d\right) }\chi \left( \alpha \beta \right)
=d\delta _{0,\beta },\qquad \chi \left( \alpha +\beta \right) =\chi \left(
\alpha \right) \chi \left( \beta \right) \,,  \label{chi_or}
\end{equation}%
similarly to the well known identities in the prime case
\begin{equation*}
\sum_{m=0}^{p-1}\omega ^{mn}=p\delta _{n,0},\qquad \omega ^{m+n}=\omega
^{m}\omega ^{n},
\end{equation*}%
where $\omega =\exp [2\pi i/p]$ is a root of unity.

In particular, for the prime dimensional case ($d=p$), the position and
momentum operators act in the standard basis $|n\rangle ,n=0,...,p-1,$ as
follows \cite{Schwinger},
\begin{equation}
Z|n\rangle =\omega ^{n}|n\rangle \ ,\qquad X|n\rangle =|n+1\rangle \ ,\quad
ZX=\omega XZ,  \label{primeZX}
\end{equation}%
where all the algebraic operations are on \textsf{mod}$\left( p\right) $, $%
n\in \mathcal{Z}_{p}$, and the rest of the operators, analogous to $%
Z_{\alpha }$ and $X_{\beta }$, can be obtained as powers of $Z$ and $X$.

It is worth noting that there is a single element of this basis
(sometimes called \textit{stabilizer state)}, labelled with the
zero element of the field, which is a common eigenstate of all
$Z_{\beta }$ with all the eigenvalues equal to the unity:
\begin{equation*}
Z_{\beta }\left\vert 0\right\rangle =\left\vert 0\right\rangle ,
\end{equation*}%
for any $\beta \in GF\left( d\right) $.

The operators (\ref{XZgf}) are related through the finite Fourier transform
operator \cite{Klimov05a}

\begin{equation}
F=\frac{1}{\sqrt{d}}\sum_{\alpha ,\beta \in GF\left( d\right) }\chi \left(
\alpha \beta \right) \left\vert \alpha \right\rangle \left\langle \beta
\right\vert ,\quad FF^{\dagger }=F^{\dagger }F=I,  \label{03_}
\end{equation}
so that

\begin{equation}
FX_{\alpha }F^{\dagger }=Z_{\alpha },  \label{ZFX}
\end{equation}%
and $F^{4}=I$ for $d=p^{n}$ where $p\neq 2$, and $F^{2}=I$ for $d=2^{n}$.
The Fourier transform offers us the possibility of introducing the \textit{%
conjugate} basis, which is related to the basis $\left\vert \alpha
\right\rangle $ as follows

\begin{equation}
\left\vert \widetilde{\alpha }\right\rangle =F\left\vert \alpha
\right\rangle ,\qquad Z_{\beta }\left\vert \widetilde{\alpha }\right\rangle
=|\widetilde{\alpha +\beta \rangle },\qquad X_{\beta }\left\vert \widetilde{%
\alpha }\right\rangle =\chi ^{\ast }\left( \alpha \beta \right) \left\vert
\widetilde{\alpha }\right\rangle ,
\end{equation}%
so that the elements of the conjugate basis are eigenvectors of the momentum
operators.

Operators $Z_{\alpha }$ and $X_{\beta }$ are particular cases of the
so-called \textit{displacement operators} which, in general, have the form
\begin{equation}
D\left( \alpha ,\beta \right) =\phi \left( \alpha ,\beta \right) Z_{\alpha
}X_{\beta },  \label{D_GF}
\end{equation}%
and the phase factor $\phi \left( \alpha ,\beta \right) $ is such that the
unitary condition

\begin{equation*}
D\left( \alpha ,\beta \right) D^{\dag }\left( \alpha ,\beta \right) =I,
\end{equation*}
is satisfied, implying that

\begin{equation}
\phi \left( \alpha ,\beta \right) \phi ^{\ast }\left( \alpha ,\beta \right)
=1.  \label{cond1}
\end{equation}%
The operational basis (\ref{D_GF}) becomes orthonormal,
\begin{equation*}
\mathrm{Tr}\left[ D\left( \alpha _{1},\beta _{1}\right) D\left( \alpha
_{2},\beta _{2}\right) \right] =d\delta _{-\alpha _{1},\alpha _{2}}\delta
_{-\beta _{1},\beta _{2}},
\end{equation*}%
or equivalently $D^{\dag }\left( \alpha ,\beta \right) =D\left( -\alpha
,-\beta \right) $, if the phase $\phi \left( \alpha ,\beta \right) $
satisfies the following condition,%
\begin{equation}
\phi \left( \alpha ,\beta \right) \phi \left( -\alpha ,-\beta \right) =\chi
\left( -\alpha \beta \right) ,  \label{phi_c2}
\end{equation}%
and for the particular case of the fields of even characteristic, $\mathrm{%
char}\left( GF\left( d\right) \right) =2,$ implies that
\begin{equation}
\phi ^{2}\left( \alpha ,\beta \right) =\chi \left( \alpha \beta \right) ,
\label{phi_c22}
\end{equation}%
which is equivalent to $D^{\dag }\left( \alpha ,\beta \right) =D\left(
\alpha ,\beta \right) $.

The displacement operators are non-Hermitian and thus, cannot be used for
mapping Hermitian operators into real phase-space functions. Nevertheless, a
desirable Hermitian kernel can be defined as the following transformation
\cite{Vourdas05,Klimov05b} of the displacement operator (\ref{D_GF}),

\begin{equation}
\Delta \left( \alpha ,\beta \right) =\frac{1}{d}\sum_{\kappa ,\lambda \in
GF\left( d\right) }\chi \left( \alpha \lambda -\beta \kappa \right) D\left(
\kappa ,\lambda \right) .  \label{k1gf}
\end{equation}%
This operator can be used for mapping operators into phase-space functions
in a self consistent way and satisfies the Stratonovich-Weyl postulates \cite%
{Strat}:

\noindent \emph{hemiticity}
\begin{equation}
\Delta \left( \alpha ,\beta \right) =\Delta ^{\dagger }(\alpha ,\beta ),
\label{K_herm}
\end{equation}%
if the condition (\ref{phi_c2}) (or (\ref{phi_c22}) in the case $p=2$) is
satisfied;

\noindent \emph{normalization}%
\begin{equation*}
\frac{1}{d}\sum_{\alpha ,\beta \in GF\left( d\right) }\Delta \left( \alpha
,\beta \right) =I;
\end{equation*}%
\emph{covariance}%
\begin{equation*}
D\left( \kappa ,\lambda \right) \Delta \left( \alpha ,\beta \right)
D^{\dagger }\left( \kappa ,\lambda \right) =\Delta \left( \alpha +\kappa
,\beta +\lambda \right) ,
\end{equation*}%
and the \emph{orthogonality relation}
\begin{equation*}
\mbox{\textrm{Tr}}\left( \Delta (\alpha ,\beta )\Delta ^{\dagger
}(\alpha ^{\prime },\beta ^{\prime })\right) =d\delta _{\alpha
,\alpha ^{\prime }}\delta _{\beta ,\beta ^{\prime }}.
\end{equation*}

Since the Stratonovich-Weyl postulates are satisfied, the symbol of an
operator $f$\ is defined in the standard way

\begin{equation}
W_{f}\left( \alpha ,\beta \right) =\mathrm{Tr}\left[ f\Delta \left( \alpha
,\beta \right) \right] ,  \label{map}
\end{equation}%
where $\mathrm{Tr}$ means the operational trace in the Hilbert space and the
inversion relation is%
\begin{equation*}
f=\frac{1}{d}\sum_{\alpha ,\beta \in GF\left( d\right) }W_{f}(\alpha ,\beta
)\Delta (\alpha ,\beta ).
\end{equation*}%
The \textit{overlap relation} has the standard form%
\begin{equation*}
\mathrm{Tr}(fg)=d\sum_{\alpha ,\beta \in GF\left( d\right) }W_{f}(\alpha
,\beta )W_{g}(\alpha ,\beta ),
\end{equation*}%
and as a particular case, the average value of the operator $f$ is
calculated as
\begin{equation*}
\left\langle f\right\rangle =d\sum_{\alpha ,\beta \in GF\left( d\right)
}W_{f}(\alpha ,\beta )W_{\rho }(\alpha ,\beta ),
\end{equation*}%
where $\rho $ is the density matrix.

In terms of the expansion coefficients of the operator $f$ in the
operational basis (\ref{D_GF})

\begin{equation}
f=\sum_{\alpha ,\beta \in GF\left( d\right) }f_{\alpha ,\beta }D\left(
\alpha ,\beta \right) ,  \label{a}
\end{equation}%
the symbol of $f$ has the form%
\begin{equation}
W_{f}(\alpha ,\beta )=\sum_{\kappa ,\lambda \in GF\left( d\right) }f_{\kappa
,\lambda }\chi ^{\ast }\left( \alpha \lambda -\beta \kappa \right) .
\label{W_f}
\end{equation}%
As some simple examples, we obtain that the symbols of the operators $%
Z_{\kappa }$ and $X_{\lambda }$ are

\begin{equation}
W_{Z_{\kappa }}\left( \alpha ,\beta \right) =\chi \left( \beta \kappa
\right) ,\quad W_{X_{\lambda }}\left( \alpha ,\beta \right) =\chi \left(
-\alpha \lambda \right) ,  \label{W_ZX}
\end{equation}%
and in the particular case of prime dimensions we get for the symbols of (%
\ref{primeZX})%
\begin{equation*}
W_{Z}\left( a,b\right) =\omega ^{b},\quad W_{X}\left( a,b\right) =\omega
^{-a},
\end{equation*}%
where $a,b\in \mathcal{Z}_{p}$. In the same way, the symbols of the basis
states $|\kappa \rangle $ and $|\widetilde{\kappa }\rangle $ are
\begin{equation*}
W_{|\kappa \rangle \langle \kappa |}\left( \alpha ,\beta \right) =\delta
_{\beta ,\kappa },\quad W_{|\widetilde{\kappa }\rangle \langle \widetilde{%
\kappa }|}\left( \alpha ,\beta \right) =\delta _{\alpha ,\kappa }.
\end{equation*}

\section{ Discrete phase space construction}

\subsection{Lines and rays}

In the discrete space $GF\left( d\right) \times GF\left( d\right) $ the
concept of line can be introduced in a similar way as in the continuous
plane case, so all the points $\left( \alpha ,\beta \right) \in GF\left(
d\right) \times GF\left( d\right) $ which satisfy the relation

\begin{equation*}
\zeta \alpha +\eta \beta =\vartheta ,\quad
\end{equation*}%
where $\zeta ,\eta ,\theta $ are some fixed elements of $GF\left( d\right) $%
, form a line. Moreover, two lines
\begin{equation}
\zeta \alpha +\eta \beta =\vartheta ,\quad \zeta ^{\prime }\alpha +\eta
^{\prime }\beta =\vartheta ^{\prime },  \label{line1}
\end{equation}%
are called parallel if they have no common points which implies that $\eta
\zeta ^{\prime }=\zeta \eta ^{\prime }$. If the lines (\ref{line1}) are not
parallel they cross each other at a single point with coordinates
\begin{equation*}
\alpha =\left( \eta \vartheta ^{\prime }-\eta ^{\prime }\vartheta \right)
\left( \zeta ^{\prime }\eta -\zeta \eta ^{\prime }\right) ^{-1},\quad \beta
=\left( \zeta \vartheta ^{\prime }-\zeta ^{\prime }\vartheta \right) \left(
\zeta \eta ^{\prime }-\zeta ^{\prime }\eta \right) ^{-1}.
\end{equation*}%
A line which passes through the origin is called \textit{ray }and its
equation has the form
\begin{equation}
\alpha =0,\quad \mbox{or}\quad \beta =\mu \alpha  \label{rays}
\end{equation}%
so that $\alpha =0$ and $\beta =0$ are the vertical and horizontal axes
correspondingly.

Each ray is characterized by the value of the "slope" $\mu $ and we denote
by $\lambda _{\mu }$ the ray which is a collection of points satisfying $%
\beta =\mu \alpha $ and by $\lambda _{\infty }$ the ray corresponding to the
vertical axis.

There are $d-1$ parallel lines to each of $d+1$ rays, so that the total
number of lines is $d(d+1)$. The collection of $d$ parallel lines is called
\textit{striation }\cite{Wootters04}.

\subsection{Displacement operators}

The displacement operators labelled with points of the phase space
belonging to the same ray commute (here we omit the phase factor):

\begin{eqnarray*}
Z_{\alpha _{1}}X_{\beta _{1}=\mu \alpha _{1}}Z_{\alpha _{2}}X_{\beta
_{2}=\mu \alpha _{2}}= \\
=\chi \left( -\mu \alpha _{1}\alpha _{2}\right) Z_{\alpha _{1}+\alpha
_{2}}X_{\mu (\alpha _{1}+\alpha _{2})}=Z_{\alpha _{2}}X_{\beta _{2}=\mu
\alpha _{2}}Z_{\alpha _{1}}X_{\beta _{1}=\mu \alpha _{1}},
\end{eqnarray*}%
and thus, have a common system of eigenvectors $\left\{ |\psi _{\nu }^{\mu
}\rangle ,\ \mu ,\nu \in GF\left( d\right) \right\} $:

\begin{equation}
Z_{\alpha }X_{\mu \alpha }|\psi _{\nu }^{\mu }\rangle =\exp (i\xi _{\mu ,\nu
})|\psi _{\nu }^{\mu }\rangle ,  \label{ZXes}
\end{equation}%
where $\mu $ is fixed and $\exp (i\xi _{\mu ,\nu })$ is the corresponding
eigenvalue, so that $\left\vert \psi _{\nu }^{0}\right\rangle \equiv |\nu
\rangle $ are eigenstates of $Z_{\alpha }$ operators (displacement operators
labelled with the points of the ray $\beta =0$ - horizontal axis) and $%
\left\vert \tilde{\psi}_{\nu }^{0}\right\rangle =F\left\vert \psi _{\nu
}^{0}\right\rangle \equiv |\tilde{\nu}\rangle $ are the eigenstates of $%
X_{\beta }$ operators (displacement operators labelled with the
points of the ray $\alpha =0$ - vertical axis).

In the simplest cases, $d=3$ and $d=4$,\ the displacement operators have the
form (up to a phase)%
\begin{equation*}
\begin{array}{cccc}
d=3 &  & d=4 &  \\
\mbox{{\small ray equation}} & \mbox{{\small displacement
operators}} &
\mbox{{\small ray equation}} & \mbox{{\small displacement operators}} \\
\beta =0 & Z,Z^{2} & \beta =0 & Z_{\theta },Z_{\theta ^{2}},Z_{\theta ^{3}}
\\
\beta =\alpha & ZX,Z^{2}X^{2} & \beta =\alpha & Z_{\theta }X_{\theta
},Z_{\theta ^{2}}X_{\theta ^{2}},Z_{\theta ^{3}}X_{\theta ^{3}} \\
\beta =2\alpha & ZX^{2},Z^{2}X^{4}=Z^{2}X & \beta =\theta \alpha & Z_{\theta
}X_{\theta ^{2}},Z_{\theta ^{2}}X_{\theta ^{3}},Z_{\theta ^{3}}X_{\theta }
\\
\alpha =0 & X,X^{2} & \beta =\theta ^{2}\alpha & Z_{\theta }X_{\theta
^{3}},Z_{\theta ^{2}}X_{\theta },Z_{\theta ^{3}}X_{\theta ^{3}} \\
&  & \alpha =0 & X_{\theta },X_{\theta ^{2}},X_{\theta ^{3}}%
\end{array}%
\end{equation*}%
where $\theta $ is the primitive element, a root of the polynomial $\theta
^{2}+\theta +1=0$.

It is easy to observe that an arbitrary displacement operator $Z_{\tau
}X_{\upsilon }$ acting on an eigenstate of the set \{$Z_{\alpha }X_{\mu
\alpha },\ \mu $ is fixed, $\alpha \in GF\left( d\right) $\} transforms it
into another eigenstate of the same set:%
\begin{equation*}
Z_{\alpha }X_{\mu \alpha }\left[ Z_{\tau }X_{\upsilon }|\psi _{\nu }^{\mu
}\rangle \right] =\exp (i\xi _{\mu ,\nu })\chi (\alpha \upsilon -\mu \alpha
\tau )Z_{\tau }X_{\upsilon }|\psi _{\nu }^{\mu }\rangle .
\end{equation*}%
It is clear that if $\upsilon =\mu \tau $ we do not generate another state
(in other words we do not change the index $\nu $). Because for an arbitrary
$\upsilon $ one can always find such $\varkappa $ that $\upsilon =\mu \tau
+\varkappa $, we can generate all the states from the set \{$|\psi _{\nu
}^{\mu }\rangle ,\mu $ is fixed\} by applying only the momentum operators $%
X_{\upsilon }$ (where $\upsilon $ runs through the whole field) to any
particular state belonging to this set. It is clear that all the eigenstates
of $X_{\beta }$ operator can be obtained by applying $Z_{\alpha }$ operator
to any particular state from the set \{$\left\vert \tilde{\psi}_{\nu
}^{0}\right\rangle $\}.

\subsection{Rotation operators}

The "rotation" operators $V_{\mu ^{\prime }}$ which transform eigenstates of
the operators associated to the ray $\beta =\mu \alpha $:

\begin{equation}
\left\{ I,Z_{\alpha _{1}}X_{\mu \alpha _{1}},Z_{\alpha _{2}}X_{\mu \alpha
_{2}},...\right\}  \label{set1}
\end{equation}%
into eigenstates of the operators labelled with points of the ray
$\beta =\left( \mu +\mu ^{\prime }\right) \alpha $:
\begin{equation}
\left\{ I,Z_{\alpha _{1}}X_{\left( \mu +\mu ^{\prime }\right) \alpha
_{1}},Z_{\alpha _{2}}X_{\left( \mu +\mu ^{\prime }\right) \alpha
_{2}},...\right\} ,\qquad  \label{set2}
\end{equation}%
are defined through the relations

\begin{equation}
V_{\mu }Z_{\alpha }V_{\mu }^{\dag }=\exp (i\varphi \left( \alpha ,\mu
\right) )Z_{\alpha }X_{\mu \alpha },\qquad \left[ V_{\mu },X_{\nu }\right]
=0,\quad V_{0}=I,  \label{s1}
\end{equation}%
for all $\mu ,\nu \in GF\left( d\right) $.

Really, being $|\psi _{\nu }^{\mu }\rangle $ (\ref{ZXes}) a state assigned
to the ray $\beta =\mu \alpha $, we obtain after simple algebra

\begin{eqnarray*}
V_{\mu ^{\prime }}Z_{\alpha }X_{\mu \alpha }|\psi _{\nu }^{\mu }\rangle
&=&V_{\mu ^{\prime }}Z_{\alpha }X_{\mu \alpha }\left( V_{\mu ^{\prime
}}^{\dag }V_{\mu ^{\prime }}\right) |\psi _{\nu }^{\mu }\rangle = \\
\exp (i\xi _{\mu ,\nu })V_{\mu ^{\prime }}|\psi _{\nu }^{\mu }\rangle
&=&\exp (i\varphi (\alpha ,\mu ^{\prime }))Z_{\alpha }X_{\left( \mu +\mu
^{\prime }\right) \alpha }V_{\mu ^{\prime }}|\psi _{\nu }^{\mu }\rangle ,
\end{eqnarray*}%
that is

\begin{equation}
Z_{\alpha }X_{\left( \mu +\mu ^{\prime }\right) \alpha }\left[ V_{\mu
^{\prime }}|\psi _{\nu }^{\mu }\rangle \right] =\exp (i(\xi _{\mu ,\nu
}-\varphi \left( \alpha ,\mu ^{\prime }\right) ))\left[ V_{\mu ^{\prime
}}|\psi _{\nu }^{\mu }\rangle \right] ,  \label{Ves}
\end{equation}%
i.e. the state $V_{\mu ^{\prime }}|\psi _{\nu }^{\mu }\rangle $ \ is an
eigenstate of the set (\ref{set2}). This means that we can interpret the
action of $V_{\mu }$ operator as a "rotation" in the discrete phase space:
\begin{equation}
\lambda _{\mu }\overset{V_{\mu ^{\prime }}}{\rightarrow }\lambda _{\mu +\mu
^{\prime }},  \label{lVl}
\end{equation}%
although it should be careful in the case of fields $GF(2^{n})$ as we will
discuss below. Observe, that one cannot reach the vertical axis by applying $%
V_{\mu }$ to any other ray.

The explicit form of $V_{\mu }$ can be found taking into account
that they are diagonal in the conjugate basis (\ref{s1}):
\begin{equation}
V_{\mu }=\sum\limits_{\kappa \in GF\left( d\right) }c_{\kappa ,\mu }|%
\widetilde{\kappa }\rangle \langle \widetilde{\kappa }|,\quad c_{0,\mu }=1.
\label{V}
\end{equation}%
Transforming the position operator $Z_{\alpha }$ with $V_{\mu }$,

\begin{equation*}
V_{\mu }Z_{\alpha }V_{\mu }^{\dag }=\sum\limits_{\kappa \in
GF\left( d\right) }c_{\kappa +\alpha ,\mu }c_{\kappa ,\mu }^{\ast
}|\widetilde{\kappa +\alpha }\rangle \langle \widetilde{\kappa }|
\end{equation*}%
and taking into account that

\begin{equation*}
Z_{\alpha }X_{\mu \alpha }=\sum\limits_{\kappa \in GF\left(
d\right) }\chi \left( -\mu \alpha \kappa \right)
|\widetilde{\kappa +\alpha }\rangle \langle \widetilde{\kappa }|,
\end{equation*}%
we obtain that the coefficients $c_{\kappa }$ satisfy the following condition

\begin{equation*}
c_{\kappa +\alpha ,\mu }c_{\kappa ,\mu }^{\ast }=\exp (i\varphi \left(
\alpha ,\mu \right) )\chi \left( -\mu \alpha \kappa \right) .
\end{equation*}%
In particular, for $\kappa =0$ we obtain
\begin{equation}
\exp (i\varphi \left( \alpha ,\mu \right) )=c_{\alpha ,\mu }c_{0,\mu }^{\ast
}=c_{\alpha ,\mu },  \label{c(mu, a)}
\end{equation}
that is

\begin{equation}
c_{\kappa +\alpha ,\mu }c_{\kappa ,\mu }^{\ast }=c_{\alpha ,\mu }\chi \left(
-\mu \alpha \kappa \right) ,  \label{ck_2}
\end{equation}%
and substituting $\alpha =0$ we get $|c_{\kappa ,\mu }|^{2}=1$, which also
follows from the unitary condition $V_{\mu }V_{\mu }^{\dagger }=V_{\mu
}^{\dagger }V_{\mu }=I$. \

It is easy to note that Eq. (\ref{ck_2}) is automatically satisfied after
the substitution%
\begin{equation*}
c_{\alpha ,\mu }\rightarrow c_{\alpha ,\mu }^{\nu }=c_{\alpha ,\mu }\chi
\left( -\alpha \nu \right) ,
\end{equation*}%
which means that all different sets of operators $V_{\mu }$ have the form%
\begin{equation}
V_{\mu ,\nu }=V_{\mu }X_{\nu },  \label{VV}
\end{equation}%
where $V_{\mu }$ is constructed using an arbitrary solution of Eq. (\ref%
{ck_2}).

\subsubsection{Fields of odd characteristic}

In the case of fields of odd characteristic, we impose an additional
restriction on the rotation operators: we demand that $V_{\mu }$ form an
Abelian group:%
\begin{equation}
V_{\mu }V_{\mu ^{\prime }}=V_{\mu +\mu ^{\prime }},  \label{V_odd}
\end{equation}%
which implies that $c_{\kappa ,\mu }c_{\kappa ,\mu ^{\prime }}=c_{\kappa
,\mu +\mu ^{\prime }}$ and in particular $c_{\kappa ,\mu }^{\ast }=c_{\kappa
,-\mu }$, leading to the relation $V_{\mu }^{\dagger }=V_{-\mu }$. In this
case the relation (\ref{lVl}) is well defined, i.e. the operator $V_{\mu }$
transforms a state associated with the ray $\lambda _{\mu }$ into a state
associated with the ray $\lambda _{\mu +\mu ^{\prime }}$. It can be shown
(see below) that the condition (\ref{V_odd}) cannot be satisfied for the
fields of even characteristic, so that this case should be considered
separately.

Then, the solution of Eq. (\ref{ck_2}) can be easily found%
\begin{equation}
c_{\kappa ,\mu }=\chi \left( -2^{-1}\kappa ^{2}\mu \right) ,  \label{ck_p}
\end{equation}%
so that
\begin{equation}
V_{\mu }=\sum\limits_{\kappa \in GF\left( d\right) }\chi \left(
-2^{-1}\kappa ^{2}\mu \right) |\widetilde{\kappa }\rangle \langle \widetilde{%
\kappa }|.  \label{vexplicit}
\end{equation}%
In the prime field case, $GF(p)$, $p\neq 2$, the whole set of rotation
operators is produced by taking powers of a single operator \cite{Vourdas}
\begin{equation*}
V=\sum\limits_{k=0}^{p-1}\omega \left( -2^{-1}k^{2}\right) |\widetilde{k}%
\rangle \langle \widetilde{k}|.
\end{equation*}%
In particular, the state $V^{m^{\prime }}|\psi _{0}^{m}\rangle $ is
associated with the ray $b=\left( m+m^{\prime }\right) a$, where the
algebraic operations are \textsf{mod}$p,$ so that
\begin{equation*}
\lambda _{0}\overset{V}{\rightarrow }\lambda _{1},\quad \underset{V^{2}}{%
\underbrace{\lambda _{0}\overset{V}{\rightarrow }\lambda _{1}\overset{V}{%
\rightarrow }\lambda _{2}}},\quad etc.
\end{equation*}

\subsubsection{Fields of even characteristic}

The situation is more complicated for fields of \textrm{char}$\left(
GF\left( d\right) \right) =2$. Really, it follows from (\ref{ck_2}) that
(substituting $\kappa =\alpha $)%
\begin{equation}
c_{\alpha ,\mu }^{2}=\chi \left( \alpha ^{2}\mu \right) .  \label{c_p2}
\end{equation}%
The solution of the above equation is not unique and thus, there is an
ambiguity in solving Eq. (\ref{ck_2}) (see Appendix B).

One of the consequences of this ambiguity is that the operators of the form (%
\ref{V}), where $c_{\kappa ,\mu }$ is a \textit{particular} (for a fixed
value of $\mu $ and $\kappa \in GF\left( 2^{n}\right) $) solution of (\ref%
{ck_2}) do not form a group. In particular, the operator $V_{\mu }^{2}$ is
not the identity operator. Really, using (\ref{c_p2}) we have

\begin{equation*}
V_{\mu }^{2}=\sum\limits_{\kappa \in GF\left( 2^{n}\right)
}c_{\kappa ,\mu
}^{2}|\widetilde{\kappa }\rangle \langle \widetilde{\kappa }%
|=\sum\limits_{\kappa \in GF\left( d\right) }\chi \left( \kappa
^{2}\mu \right) |\widetilde{\kappa }\rangle \langle
\widetilde{\kappa }|,
\end{equation*}%
which due to the property $\mathrm{tr}\alpha =\mathrm{tr}\alpha ^{2}$, $%
\alpha \in GF\left( 2^{n}\right) $, can be transformed into

\begin{equation}
V_{\mu }^{2}=\sum\limits_{\kappa \in GF\left( 2^{n}\right) }\chi
\left(
\kappa \mu ^{2^{n-1}}\right) |\widetilde{\kappa }\rangle \langle \widetilde{%
\kappa }|=X_{\mu ^{2^{n-1}}},  \label{V2_2}
\end{equation}%
where the relation $\kappa ^{2}\mu =\left( \kappa \mu ^{2^{n-1}}\right) ^{2}$
has been used.

It also follows from (\ref{V2_2}) that inside the set $\{V_{\mu },\mu \in
GF\left( 2^{n}\right) \}$ an inverse operator to a given $V_{\mu }$ from
this set does not exist. To find the inverse operator to some $V_{\mu }$ we
have to extend the set $\{V_{\mu },\ \mu \in GF\left( 2^{n}\right) \}$ to
the whole collection of all the possible rotation operators defined in (\ref%
{VV}), i.e. to the set $\{V_{\mu ,\nu },\ \mu ,\nu \in GF\left( 2^{n}\right)
\}$. Then, it is easy to conclude from (\ref{V2_2}) that
\begin{equation*}
\left( V_{\mu ,\nu }\right) ^{-1}=V_{\mu ,\mu ^{2^{n-1}}+\nu },
\end{equation*}%
which implies the following relation between $c_{\alpha ,\mu }$:
\begin{equation}
c_{\alpha ,\mu }^{\ast }=\chi \left( \alpha \mu ^{2^{n-1}}\right) c_{\alpha
,\mu }.  \label{c_conj}
\end{equation}

We will fix the operator $V_{\mu ,\nu =0}$ is such a way (see Appendix B)
that the coefficients $c_{\kappa ,\mu }$, corresponding to the basis
elements of the field, $\kappa =\sigma _{1},...,\sigma _{n}$, are chosen
positive, so that we have
\begin{eqnarray}
c_{\kappa ,\mu } &=&\chi \left( \mu \sum_{i=1}^{n-1}k_{i}\sigma
_{i}\sum_{j=i+1}^{n}k_{j}\sigma _{j}\right) \Pi _{l=1}^{n}\sqrt{\chi \left(
k_{l}^{2}\sigma _{l}^{2}\mu \right) },  \label{c0} \\
\kappa &=&\sum_{i=1}^{n}k_{i}\sigma _{i},\quad k_{i}\in
\mathcal{Z}_{2},
\end{eqnarray}%
where $\{\sigma _{i},i=1,..n\}$ are the elements of the basis and the
principal branch of the square root in \ (\ref{c0}) is chosen. All the other
possible rotating operators can be obtained according to (\ref{VV}).

The whole set of operators $\{V_{\mu ,\nu },\ \mu ,\nu \in GF\left(
2^{n}\right) \}$ form a group. Really, let us pick two operators $V_{\mu }$
and $V_{\mu ^{\prime }}$ constructed according to (\ref{c0}). Using the
properties (\ref{s1}) and (\ref{c(mu, a)}) we obtain%
\begin{equation}
V_{\mu }V_{\mu ^{\prime }}Z_{\alpha }V_{\mu ^{\prime }}^{\dag }V_{\mu
}^{\dag }=c_{\alpha ,\mu }c_{\alpha ,\mu ^{\prime }}Z_{\alpha }X_{\left( \mu
+\mu ^{\prime }\right) \alpha }.
\end{equation}%
On the other hand we have%
\begin{equation*}
V_{\mu +\mu ^{\prime }}Z_{\alpha }V_{\mu +\mu ^{\prime }}^{\dag }=c_{\alpha
,\mu +\mu ^{\prime }}Z_{\alpha }X_{\left( \mu +\mu ^{\prime }\right) \alpha
},
\end{equation*}%
which suggests that (recall that $|c_{\alpha ,\mu }|=1$)
\begin{equation}
c_{\alpha ,\mu }c_{\alpha ,\mu ^{\prime }}=\exp \left( if(\alpha ,\mu ,\mu
^{\prime })\right) c_{\alpha ,\mu +\mu ^{\prime }},  \label{cc_mu}
\end{equation}%
where $f(\alpha ,\mu ,\mu ^{\prime })=f(\alpha ,\mu ^{\prime },\mu )$ is a
real function of $\alpha ,\mu $, and $\mu ^{\prime }$. Let us note that due
to the property (\ref{ck_2}), the function $f(\alpha ,\mu ,\mu ^{\prime })$
linearly depends on the parameter $\alpha $, in the sense that $f(\alpha
+\beta ,\mu ,\mu ^{\prime })=f(\alpha ,\mu ,\mu ^{\prime })+f(\beta ,\mu
,\mu ^{\prime })$. The complex conjugate of (\ref{cc_mu}) together with (\ref%
{c_conj}) leads to the condition $\exp \left( if(\alpha ,\mu ,\mu ^{\prime
})\right) =\exp \left( -if(\alpha ,\mu ,\mu ^{\prime })\right) =\pm 1\in
\mathcal{Z}_{2}$. Because every linear map from $\alpha \in GF(p^{n})$ to $%
GF(p)\simeq \mathcal{Z}_{p}$ is of the form of a character \cite{FF}: $%
\alpha \rightarrow \chi \left( \alpha \beta \right) ,\beta \in GF(p^{n})$,
the function $\exp \left( if(\alpha ,\mu ,\mu ^{\prime })\right) $ can be
represented as%
\begin{equation*}
\exp \left( if(\alpha ,\mu ,\mu ^{\prime })\right) =\chi \left( \alpha f(\mu
,\mu ^{\prime })\right) .
\end{equation*}%
It follows from (\ref{cc_mu}) and from the above equation that there exists $%
\nu =f(\mu ,\mu ^{\prime })$ (see Appendix C) so that

\begin{equation}
V_{\mu }V_{\mu ^{\prime }}Z_{\alpha }V_{\mu \prime }^{\dag }V_{\mu }^{\dag
}=X_{\nu }V_{\mu +\mu ^{\prime }}Z_{\alpha }V_{\mu +\mu ^{\prime }}^{\dag
}X_{\nu }^{\dag }=c_{\alpha ,\mu +\mu ^{\prime }}\chi \left( \alpha \nu
\right) Z_{\alpha }X_{(\mu +\mu ^{\prime })\alpha },  \label{VX}
\end{equation}%
which immediately leads to the relation

\begin{equation*}
V_{\mu }V_{\mu ^{\prime }}=V_{\mu +\mu ^{\prime }}X_{f(\mu ,\mu ^{\prime })}.
\end{equation*}%
Finally, in general we have
\begin{equation*}
V_{\mu ,\nu }V_{\mu ^{\prime },\nu ^{\prime }}=V_{\mu +\mu ^{\prime },\nu
+\nu ^{\prime }+f(\mu ,\mu ^{\prime })}.
\end{equation*}

\subsection{Phase space construction}

We can associate the lines in the discrete phase space with states in the
Hilbert space according to the following construction:

1. The eigenstate of \{$Z_{\alpha }$\} operators with all eigenvalues equal
to $1$, $|0\rangle =|\psi _{0}^{0}\rangle $, (note that such state is
unique) is associated with the horizontal axis, $\beta =0$. It is worth
noting that such an association is arbitrary and in some sense fixes a
definite quantum net \cite{Wootters04}.

2. All the other states of the "first" striation are obtained by applying
the displacement operator $X_{\nu }$ to $|0\rangle $, so that the state $%
|\psi _{\nu }^{0}\rangle =X_{\nu }|0\rangle $ is associated with the
horizontal line which crosses the vertical axis at the point $(0,\nu )$,
i.e. with the line $\beta =\nu $. The states $|\psi _{\nu }^{0}\rangle $ are
eigenstates of the set \{$Z_{\alpha }$\}:%
\begin{equation}
Z_{\alpha }|\psi _{\nu }^{0}\rangle =\chi (\alpha \nu )X_{\nu }Z_{\alpha
}|0\rangle =\chi (\alpha \nu )|\psi _{\nu }^{0}\rangle ,  \label{Z_psc}
\end{equation}%
and form an orthonormal basis $\langle \psi _{\nu }^{0}|\psi _{\nu ^{\prime
}}^{0}\rangle =\delta _{\nu \nu ^{\prime }}.$

3. All the other striations are constructed as follows: First we apply the
rotation operator $V_{\mu }$ to the state $|0\rangle $ and the obtained
state $|\psi _{0}^{\mu }\rangle =V_{\mu }|0\rangle $ is associated to the
ray $\beta =\mu \alpha $. The state $|\psi _{0}^{\mu }\rangle $ is an
eigenstate of the set \{$Z_{\alpha }X_{\mu \alpha }$\} according to (\ref%
{Ves}). It is worth noting that different sets of rotation operators can be
chosen, which leads to different associations between states and lines (see
discussion below).

4. All the other states of the $\mu $-th striation are obtained by applying
the operator $X_{\nu }$ to the state $|\psi _{0}^{\mu }\rangle $:%
\begin{equation}
|\psi _{\nu }^{\mu }\rangle =X_{\nu }|\psi _{0}^{\mu }\rangle .  \label{VXes}
\end{equation}%
The states $|\psi _{\nu }^{\mu }\rangle $ are eigenstates of the set \{$%
Z_{\alpha }X_{\mu \alpha }$\}:%
\begin{eqnarray*}
Z_{\alpha }X_{\mu \alpha }|\psi _{\nu }^{\mu }\rangle &=&\exp (-i\varphi
\left( \alpha ,\mu \right) )V_{\mu }Z_{\alpha }V_{\mu }^{\dagger }X_{\nu
}V_{\mu }|0\rangle = \\
\exp (-i\varphi \left( \alpha ,\mu \right) )V_{\mu }Z_{\alpha }X_{\nu
}|0\rangle &=&\chi (\alpha \nu )\exp (-i\varphi \left( \alpha ,\mu \right)
)X_{\nu }V_{\mu }|0\rangle \\
&=&\chi (\alpha \nu )\exp (-i\varphi \left( \alpha ,\mu \right) )|\psi _{\nu
}^{\mu }\rangle ,
\end{eqnarray*}%
and are associated to the lines $\beta =\mu \alpha +\nu $. The phase $\exp
(-i\varphi \left( \alpha ,\mu \right) )$ in the above equation is defined in
(\ref{s1}).

Note that $|\psi _{\nu }^{\mu }\rangle =V_{\mu ,\nu }|0\rangle ,$ where $%
V_{\mu ,\nu }$ is defined in (\ref{VV}). As it was shown in \cite{Klimov05a}
the states (\ref{VXes}) form mutually unbiased bases ($MUB$s) \cite%
{WoottersU}:%
\begin{equation*}
|\langle \psi _{\nu }^{\mu }|\psi _{\nu ^{\prime }}^{\mu ^{\prime }}\rangle
|^{2}=\frac{1}{d},\quad \mu \neq \mu ^{\prime },
\end{equation*}%
which have been extensively discussed from different points of view in
recent papers \cite{MUB1, MUB2, Klimov05a, Durt05}. In particular, it is
known \cite{Wootters04, Rubin05} that different $MUB$s can be associated
with different striations.

In our approach we note that the inner product of two states
associated to lines can be rewritten as follows
\begin{equation}
\langle \psi _{\nu }^{\mu }|\psi _{\nu ^{\prime }}^{\mu ^{\prime }}\rangle
=\langle 0|V_{\mu }^{\dagger }X_{\nu }^{\dagger }X_{\nu ^{\prime }}V_{\mu
^{\prime }}|0\rangle =\langle 0|X_{\nu ^{\prime }-\nu }V_{\mu ^{\prime }-\mu
}|0\rangle =\sum_{\kappa \in GF\left( d\right) }c_{\kappa ,\mu ^{\prime
}-\mu }\langle \nu -\nu ^{\prime }|\widetilde{\kappa }\rangle \langle
\widetilde{\kappa }|0\rangle ,  \label{opsaverage}
\end{equation}%
so that, taking into account $\langle \nu |\widetilde{\kappa }\rangle =\chi
(\kappa \nu )/d^{1/2}$, we obtain
\begin{equation*}
\langle \psi _{\nu }^{\mu }|\psi _{\nu ^{\prime }}^{\mu ^{\prime }}\rangle =%
\frac{1}{d}\sum_{\kappa \in GF\left( d\right) }c_{\kappa ,\mu ^{\prime }-\mu
}\chi (\kappa (\nu -\nu ^{\prime }))=\frac{1}{d}\Psi _{\nu ,\nu ^{\prime
}}^{\mu ,\mu ^{\prime }}.
\end{equation*}%
Then, we get%
\begin{equation*}
|\Psi _{\nu ,\nu ^{\prime }}^{\mu ,\mu ^{\prime }}|^{2}=\sum_{\kappa ,\kappa
^{\prime }\in GF\left( d\right) }c_{\kappa ,\mu ^{\prime }-\mu }c_{\kappa
^{\prime },\mu ^{\prime }-\mu }^{\ast }\chi ((\kappa -\kappa ^{\prime })(\nu
-\nu ^{\prime })),
\end{equation*}%
so that after changing the index $\kappa -\kappa ^{\prime }=\lambda $, and
making use the relation (\ref{ck_2}){} we immediately obtain

\begin{eqnarray*}
|\Psi _{\nu ,\nu ^{\prime }}^{\mu ,\mu ^{\prime }}|^{2} &=&\sum_{\kappa
^{\prime },\lambda \in GF\left( d\right) }c_{\lambda ,\mu ^{\prime }-\mu
}\chi (-(\mu ^{\prime }-\mu )\lambda \kappa ^{\prime })\chi (\lambda (\nu
-\nu ^{\prime })) \\
&=&\left\{
\begin{array}{cc}
d^{2}\delta _{\nu ,\nu ^{\prime }} & \mu =\mu ^{\prime } \\
d\sum_{l=0}^{p-1}c_{\lambda ,\mu ^{\prime }-\mu }\chi (\lambda (\nu -\nu
^{\prime }))\delta _{0,\lambda }=d & \mu \neq \mu ^{\prime }%
\end{array}%
\right. .
\end{eqnarray*}%
The uppermost relation corresponds to the inner product of states belonging
to the same striation, i.e. to the same basis, and from the other one can
observe that different striations correspond to different $MUB$s.

5. The state associated with the vertical axis $\beta $ is obtained from $%
|0\rangle $ by applying the Fourier transform operator:%
\begin{equation}
|\tilde{0}\rangle =F|0\rangle .  \label{Xes}
\end{equation}%
The state (\ref{Xes}) is an eigenstate of the set \{$X_{\beta },\ \beta \in
GF(d)$\} with all eigenvalues equal to unity: $X_{\beta }|\tilde{0}\rangle =|%
\tilde{0}\rangle $. The vertical lines, parallel to the axis $\beta $,
crossing the axis $\alpha $ at the points $\alpha =\nu $ are associated with
the states $|\tilde{\psi}_{\nu }^{0}\rangle =Z_{\nu }|\tilde{0}\rangle ,$
which are eigenstates of the set \{$X_{\beta }$\}:%
\begin{equation*}
X_{\beta }|\tilde{\psi}_{\nu }^{0}\rangle =\chi (-\beta \nu )Z_{\nu
}X_{\beta }|\tilde{0}\rangle =\chi (-\beta \nu )|\tilde{\psi}_{\nu
}^{0}\rangle .
\end{equation*}%
It is clear that
\begin{equation*}
|\langle \psi _{\nu }^{\mu }|\tilde{\psi}_{\nu ^{\prime }}^{0}\rangle |^{2}=%
\frac{1}{d}.
\end{equation*}

For the fields of odd characteristics we can always choose the set of
rotation operators in such a way that they form a group, which in particular
means that a consecutive application of any $V_{\mu }$ from this set returns
us to the initial state: $V_{\mu }^{p}=I$. This does not hold for the fields
of even characteristic due to Eq.(\ref{V2_2}), which can be interpreted as
follows: the first application of the rotation operator $V_{\mu }$
transforms a state associated with the ray of some striation to a state
corresponding to a ray from another striation. The second application of $%
V_{\mu }$ does not return to the initial state (as it could be expected from
(\ref{lVl})), but to a state which belongs to the original striation and
displaced by $X_{\mu ^{2^{n-1}}}$. So that, from the geometrical point of
view the operator $V_{\mu }^{2}$ for $GF(2^{n})$ transforms a ray from some
striation to a parallel line from the same striation.

\subsection{The phase of the displacement operator}

The phase $\varphi \left( \alpha ,\mu \right) $ which appears in (\ref{s1})
is intimately related to the phase $\phi \left( \alpha ,\beta \right) $ of
the displacement operator (\ref{D_GF}). Let us impose the natural condition
that the sum of the Wigner function along the line $\beta =\mu \alpha +\nu $
is equal to the average value of the density matrix over the state $%
\left\vert \psi _{\nu }^{\mu }\right\rangle $ associated with that line \cite%
{Wootters}:

\begin{equation}
\frac{1}{d}\sum\limits_{\alpha ,\beta \in GF\left( d\right)
}W\left( \alpha ,\beta \right) \delta _{\beta ,\mu \alpha +\nu
}=\left\langle \psi _{\nu }^{\mu }\right\vert \rho \left\vert \psi
_{\nu }^{\mu }\right\rangle , \label{W_rho_1}
\end{equation}%
where

\begin{equation}
W\left( \alpha ,\beta \right) =\mathrm{Tr}\left[ \rho \Delta \left( \alpha
,\beta \right) \right] .  \label{W}
\end{equation}%
According to the general construction, the state $\left\vert \psi _{\nu
}^{\mu }\right\rangle $ is obtained by application of the operator $V_{\mu
,\nu }$ to the state $\left\vert 0\right\rangle $: $\left\vert \psi _{\nu
}^{\mu }\right\rangle =V_{\mu ,\nu }\left\vert 0\right\rangle $.\ After
simple algebra we transform the left-hand side of (\ref{W_rho_1}) as follows

\begin{equation}
\frac{1}{d}\sum\limits_{\alpha ,\beta \in GF\left( d\right)
}W\left( \alpha
,\beta \right) \delta _{\beta ,\mu \alpha +\nu }=\frac{1}{d}%
\sum\limits_{\alpha \in GF\left( d\right) }W\left( \alpha ,\mu
\alpha +\nu \right)  \label{line1a}
\end{equation}%
\begin{equation*}
=\frac{1}{d}\sum\limits_{\kappa ,\lambda \in GF\left( d\right)
}\rho _{\kappa ,\lambda }\phi \left( \mu ^{-1}\left( -\kappa
+\lambda \right) ,-\kappa +\lambda \right) \chi \left( \mu
^{-1}\left( \kappa +\lambda \right) \left( \lambda +\nu \right)
\right) ,
\end{equation*}%
where $\rho =\sum_{\kappa ,\lambda }\rho _{\kappa ,\lambda }|\kappa \rangle
\langle \lambda |$, meanwhile the right-hand part is converted into%
\begin{eqnarray*}
\left\langle \psi _{\nu }^{\mu }\right\vert \rho \left\vert \psi _{\nu
}^{\mu }\right\rangle &=&\left\langle 0\right\vert V_{\mu }^{\dag }X_{\nu
}^{\dagger }\rho X_{\nu }V_{\mu }\left\vert 0\right\rangle \\
&=&\frac{1}{d^{2}}\sum\limits_{\kappa ,\kappa ^{\prime },\tau
,\upsilon }\rho _{\tau ,\upsilon }c_{\kappa ,\mu }^{\ast
}c_{\kappa ^{\prime },\mu }\chi \left( -\kappa \left( \tau -\nu
\right) +\kappa ^{\prime }\left( \upsilon -\nu \right) \right) .
\end{eqnarray*}%
Taking into account the relation (\ref{ck_2}), the above equation can be
rewritten as follows

\begin{equation}
\left\langle \psi _{\nu }^{\mu }\right\vert \rho \left\vert \psi
_{\nu }^{\mu }\right\rangle =\frac{1}{d}\sum\limits_{\kappa
,\lambda }\rho _{\kappa ,\lambda }c_{\mu ^{-1}\left( -\kappa
+\lambda \right) ,\mu }\chi \left( \mu ^{-1}\left( \kappa +\lambda
\right) \left( \lambda +\nu \right) \right) .  \label{line2}
\end{equation}%
Comparing (\ref{line1a}) and (\ref{line2}) we observe that $\phi \left( \mu
^{-1}\left( -\kappa +\lambda \right) ,-\kappa +\lambda \right) =c_{\mu
^{-1}\left( -\kappa +\lambda \right) ,\mu },$ or in a compact form

\begin{equation}
\phi \left( \tau ,\upsilon \right) =c_{\tau ,\tau ^{-1}\upsilon }.
\label{phi_gen}
\end{equation}%
Also, we impose the conditions $\phi \left( \tau ,0\right) =\phi \left(
0,\upsilon \right) =1$, which means that the displacements along the axes $%
\alpha $ and $\beta $, which are performed by applying $Z_{\kappa }$ and $%
X_{\lambda }$ operators correspondingly, do not generate any phase.

This immediately leads that for fields of odd characteristic
\begin{equation}
\phi \left( \tau ,\upsilon \right) =\chi \left( -2^{-1}{}\tau \upsilon
\right) ,  \label{phi_odd}
\end{equation}%
and the condition (\ref{phi_c2}) is automatically satisfied. Thus,
the displacement operator in this case has the form
\begin{equation}
D\left( \alpha ,\beta \right) =\chi \left( -2^{-1}\alpha \beta \right)
Z_{\alpha }X_{\beta }  \label{D_odd}
\end{equation}%
so that the unitary condition $D^{\dag }\left( \alpha ,\beta \right)
=D\left( -\alpha ,-\beta \right) $ is satisfied and the kernel (\ref{k1gf})
can be represented in the familiar form
\begin{equation*}
\Delta \left( \alpha ,\beta \right) =D\left( \alpha ,\beta \right) PD^{\dag
}\left( \alpha ,\beta \right) ,
\end{equation*}%
where
\begin{equation*}
P=\frac{1}{d}\sum_{\alpha ,\beta \in GF\left( d\right) }D\left( \alpha
,\beta \right) ,\qquad P\left\vert \alpha \right\rangle =\left\vert -\alpha
\right\rangle ,
\end{equation*}%
is the \textit{parity operator}.

In general, for a density matrix defined in the standard basis $\left\vert
\alpha \right\rangle $ as

\begin{equation}
\rho =\sum\limits_{\mu ,\nu \in GF\left( d\right) }\rho _{\mu ,\nu
}\left\vert \mu \right\rangle \left\langle \nu \right\vert ,
\label{rho_mn}
\end{equation}%
the Wigner function takes the form%
\begin{equation}
W_{\rho }\left( \alpha ,\beta \right)
=\frac{1}{d}\sum\limits_{\gamma ,\mu ,\nu }\chi \left( \gamma
\left( \nu -\beta \right) +\alpha \left( \nu -\mu \right) \right)
\phi \left( \gamma ,\nu -\mu \right) \rho _{\mu ,\nu },
\label{W_rho1}
\end{equation}%
which still can be simplified (summed over $\gamma $) for the fields of odd
characteristics taking into account the explicit expression (\ref{phi_odd})
for the phase $\phi $. Nevertheless, for the fields of even characteristic, $%
d=2^{n}$, there is a freedom in the election of the phase $\phi $, which
takes values $\pm 1,\pm i$, related to different possibilities of choosing
the rotation operators $V_{\mu ,\nu }=V_{\mu }X_{\nu }$ for the phase space
construction. Once fixed the set $\{V_{\mu ,\nu }\}$ of rotation operators
we can find the phase $\phi $ from (\ref{phi_gen}) and thus, construct the
Wigner function (\ref{W_rho1}).

\subsection{Non-uniqueness of the Wigner functions}

It is worth noting that the Wigner function constructed using different
distributions of signs in $V_{\mu }$ operators are not trivially related to
each other. This difference is of fundamental significance for fields of
even characteristic (because there is no natural choice of the set of
rotation operators), although similar considerations can also be taken into
account for fields of odd characteristic (if different to (\ref{vexplicit})
set of rotation operators is used for the phase-space construction).

In the rest of this section we will focus on fields of even characteristic.
Let us fix operators $V_{\mu }$ according to (\ref{c0}) and choose some
other set, so that $V_{\mu ,h(\mu )}=V_{\mu }X_{h(\mu )}$, where $h(\mu )$
is an arbitrary function satisfying the conditions: a) $h(0)=0$, which
basically means that even after changing the rotation operators, the
displacement operators along the axes $\alpha $ and $\beta $ do not have
phases, $D\left( \alpha ,0\right) =Z_{\alpha }$, $D\left( 0,\beta \right)
=X_{\beta }$; b) the non- singularity: $\alpha h\left( \alpha ^{-1}\beta
\right) =0,$ if $\alpha =0$ for any $\beta \in GF\left( 2^{n}\right) $,
which implies that the coefficient $c_{0,\mu }$ in (\ref{V}) is fixed, $%
c_{0,\mu }=1$, for all $V_{\mu }$. The simplest example of such
function can be given in the case when rotation operators are
labelled with powers of a primitive element $\sigma $: $\{V_{\mu
},\mu \in GF\left( 2^{n}\right)
\}=\{V_{\sigma ^{k}},k=1,..,2^{n}-1\}$, then $h(\sigma ^{k})=\sigma ^{m(k)}$%
, where $m$ is a natural number which depends on the value of $k$.

The Wigner function constructed using the new rotation operators $V_{\mu
,\nu }$ has the form%
\begin{equation*}
W_{\rho }^{\prime }\left( \alpha ,\beta \right) =\mathrm{Tr}\left[ \rho
\Delta ^{\prime }\left( \alpha ,\beta \right) \right] ,\quad \Delta ^{\prime
}\left( \alpha ,\beta \right) =\frac{1}{2^{n}}\sum_{\kappa ,\lambda \in
GF\left( 2^{n}\right) }\chi \left( \alpha \lambda -\beta \kappa \right)
D^{\prime }\left( \kappa ,\lambda \right) ,
\end{equation*}%
where%
\begin{equation*}
D^{\prime }\left( \kappa ,\lambda \right) =c_{\kappa ,\kappa ^{-1}\lambda
}^{\prime }Z_{\kappa }X_{\lambda },
\end{equation*}%
and $c_{\kappa ,\xi }^{\prime }$ are the matrix elements (\ref{V}) of $%
V_{\mu ,h(\mu )}$ in the conjugate basis $|\tilde{\alpha}\rangle $, so that
\begin{equation*}
c_{\kappa ,\kappa ^{-1}\lambda }^{\prime }=\chi \left( \kappa h\left( \kappa
^{-1}\lambda \right) \right) c_{\kappa ,\kappa ^{-1}\lambda }.
\end{equation*}%
Then, we have
\begin{eqnarray*}
D^{\prime }\left( \kappa ,\lambda \right) &=&\chi \left( \kappa h\left(
\kappa ^{-1}\lambda \right) \right) c_{\kappa ,\kappa ^{-1}\lambda
}Z_{\kappa }X_{\lambda }=c_{\kappa ,\kappa ^{-1}\lambda }X_{h\left( \kappa
^{-1}\lambda \right) }Z_{\kappa }X_{h\left( \kappa ^{-1}\lambda \right)
}X_{\lambda } \\
&=&X_{h\left( \kappa ^{-1}\lambda \right) }D\left( \kappa ,\lambda \right)
X_{h\left( \kappa ^{-1}\lambda \right) }.
\end{eqnarray*}%
After some simple algebra we obtain the "new" Wigner function $W_{\rho
}^{\prime }\left( \alpha ,\beta \right) $ in terms of coefficients of the
expansion (\ref{rho_mn})%
\begin{eqnarray}
W_{\rho }^{\prime }\left( \alpha ,\beta \right) = \frac{1}{2^{n}}%
\sum\limits_{\mu ,\nu ,\gamma}\rho _{\mu ,\nu }\chi \left( \alpha
\left( \nu -\mu \right) -\beta \gamma + \gamma h\left( \gamma
^{-1}\left( \nu -\mu \right) \right) +\gamma \nu \right) \phi
\left( \gamma ,\nu -\mu \right), \label{W_p1}
\end{eqnarray}%
which is related to the "old" Wigner function (constructed with $V_{\mu }$
operators) as follows
\begin{eqnarray}
W_{\rho }^{\prime }\left( \alpha ,\beta \right) &=&\frac{1}{2^{n}}%
\sum\limits_{\underset{\gamma \neq 0}{\gamma ^{\prime }}\in
GF\left( 2^{n}\right) }W_{\rho }\left( \alpha +\gamma h\left(
\gamma ^{-1}\right)
+\gamma \gamma ^{\prime },\beta +\gamma ^{\prime }\right)  \label{W_p2} \\
&&+\frac{1}{2^{n}}\sum\limits_{\beta \in GF\left( 2^{n}\right)
}W_{\rho }\left( \alpha ,\beta \right)
+\frac{1}{2^{n}}\sum\limits_{\alpha \in GF\left( 2^{n}\right)
}W_{\rho }\left( \alpha ,\beta \right) -1.
\end{eqnarray}%
Note, that
\begin{equation*}
p(\alpha )=\sum\limits_{\beta \in GF\left( 2^{n}\right) }W_{\rho
}\left( \alpha ,\beta \right) ,\quad \widetilde{p}(\beta
)=\sum\limits_{\alpha \in GF\left( 2^{n}\right) }W_{\rho }\left(
\alpha ,\beta \right) ,
\end{equation*}%
are the marginal probabilities to detect the system at the states $|\alpha
\rangle $ and $|\tilde{\beta}\rangle $ correspondingly.

It is clear that the sum of the "new" Wigner function over a line $\beta
=\mu \alpha +\nu $ gives the same result as the sum of the "old" Wigner
function over the points of the shifted line $\beta =\mu \alpha +\nu
+h\left( \mu \right) $:

\begin{equation*}
\sum\limits_{\alpha ,\beta \in GF\left( 2^{n}\right) }W^{\prime
}\left( \alpha ,\beta \right) \delta _{\beta =\mu \alpha +\nu
}=\sum\limits_{\alpha ,\beta \in GF\left( 2^{n}\right) }W\left(
\alpha ,\beta \right) \delta _{\beta =\mu \alpha +\nu +h\left( \mu
\right) }.
\end{equation*}

For a particular choice $h\left( \mu \right) =\kappa $ - $const$, $h(0)=0,$
we find that the new Wigner function has the form
\begin{equation}
W_{\rho }^{\prime }\left( \alpha ,\beta \right) =W_{\rho }\left( \alpha
,\beta +\kappa \right) -\rho _{\beta +\kappa ,\beta +\kappa }+\rho _{\beta
,\beta }.  \label{Wbk}
\end{equation}%
For a more complicated case $h\left( \xi \right) =\kappa \xi $ the Wigner
function acquires the following form%
\begin{equation*}
W_{\rho }^{\prime }\left( \alpha ,\beta \right) =W_{\rho }\left( \alpha
+\kappa ,\beta \right) +\frac{1}{2^{n}}p(\alpha )-\frac{1}{2^{n}}p(\alpha
+\kappa ).
\end{equation*}

Using the relation (\ref{W}) we can find the Wigner function for the state (%
\ref{VXes}) corresponding to the line $\beta =\mu \alpha +\nu $. First of
all, using (\ref{V}) and the property $\langle \tilde{\kappa}|\lambda
\rangle =\chi (-\kappa \lambda )/d^{1/2}$ we rewrite the state $|\psi _{\nu
}^{\mu }\rangle $ as follows%
\begin{equation*}
|\psi _{\nu }^{\mu }\rangle =\frac{1}{\sqrt{d}}\sum_{\kappa \in GF\left(
d\right) }c_{\kappa ,\mu }\chi (-\kappa \nu )|\tilde{\kappa}\rangle =\frac{1%
}{d}\sum_{\kappa ,\lambda \in GF\left( d\right) }c_{\kappa ,\mu }\chi
(\kappa \left( \lambda -\nu \right) )|\lambda \rangle .
\end{equation*}%
Then, from (\ref{k1gf}) and (\ref{W}) we obtain after long but
straightforward algebra%
\begin{equation*}
W_{|\psi _{\nu }^{\mu }\rangle }(\alpha ,\beta )=\langle \psi _{\nu }^{\mu
}|\Delta (\alpha ,\beta )|\psi _{\nu }^{\mu }\rangle =\delta _{\beta ,\mu
\alpha +\nu },
\end{equation*}%
for fields of both odd and even characteristic independently on the choice
of the rotation operator in the last case.

In general, the Wigner function of any state of the form (\ref{rho_mn}) with
$\rho _{\mu ,\nu }=q_{\mu }\delta _{\mu ,\nu }$ does not depend on the sign
choice (and thus, is uniquely defined) as it can be observed directly from (%
\ref{W_p1}).

As a nontrivial example of essentially different Wigner functions which can
be associated to the same state let us consider a particular case of pure
states with real coefficients, i.e. $\rho _{\mu ,\nu }=\rho _{\nu ,\mu }$.
For this class of states the Wigner function (\ref{W_p1}) can be rewritten
in the following explicitly symmetric form%
\begin{eqnarray*}
W_{\rho }^{\prime }\left( \alpha ,\beta \right) =\frac{1}{2^{n+1}}%
\sum\limits_{\underset{\gamma \neq 0}{\mu \neq \nu }\in GF\left(
2^{n}\right) }\rho _{\mu ,\nu }\chi \left( \alpha \left( \nu +\mu
\right) +\beta \gamma +\gamma h\left( \gamma ^{-1}\left( \nu +\mu
\right) \right)
\right)  \\
\times \phi \left( \gamma ,\nu +\mu \right)
\left[ \chi \left( \gamma \nu \right) +\chi \left( \gamma \mu \right) %
\right] +\frac{1}{2^{n}}\sum\limits_{\mu \neq \nu \in GF\left(
2^{n}\right) }\rho _{\mu ,\nu }\chi \left( \alpha \left( \nu -\mu
\right) \right) +\rho _{\beta ,\beta }.
\end{eqnarray*}%
Now, we observe from the above equation that\textbf{\ }if\textbf{\ }the
factor $\left[ \chi \left( \gamma \nu \right) +\chi \left( \gamma \mu
\right) \right] $ is zero, the Wigner function obviously does not depend on
the choice of the function $h\left( \nu \right) $. Thus, two Wigner
functions corresponding to functions $h_{1}(\nu )$ and $h_{2}\left( \nu
\right) $ are the same if simultaneously $\chi \left( \gamma h_{1}\left(
\gamma ^{-1}\left( \nu +\mu \right) \right) \right) =\chi \left( \gamma
h_{2}\left( \gamma ^{-1}\left( \nu +\mu \right) \right) \right) $ and $\chi
\left( \gamma \nu \right) +\chi \left( \gamma \mu \right) \neq 0$ or, in
other words,
\begin{eqnarray}
\mathrm{tr}\left( \gamma \left[ h_{1}\left( \gamma ^{-1}\left( \nu +\mu
\right) \right) +h_{2}\left( \gamma ^{-1}\left( \nu +\mu \right) \right) %
\right] \right) &=&0,  \label{nw1} \\
\mathrm{tr}\left( \gamma \left[ \mu +\nu \right] \right) &=&0,  \label{nw2}
\end{eqnarray}%
where $\gamma \neq 0$ and $\mu \neq \nu $.

For instance, consider the state $\left\vert \psi \right\rangle =\left(
\left\vert 0\right\rangle +\left\vert \sigma ^{3}\right\rangle \right) /%
\sqrt{2}$ in the case $GF\left( 2^{2}\right) $ and fix the irreducible
polynomial as $\sigma ^{2}+\sigma +1=0$. The indices $\mu $ and $\nu $ in (%
\ref{nw1})-(\ref{nw2}) take on values $0$ and $\sigma ^{3}$, so that $\mu
+\nu =\sigma ^{3}$, and thus, it follows from (\ref{nw2}) that the only
admissible value of $\gamma $ is $\gamma =\sigma ^{3}$. Then, the condition (%
\ref{nw1}) leads to the following equation for the functions $h_{1}(\nu )$
and $h_{2}\left( \nu \right) $:

\begin{equation}
h_{1}\left( \sigma ^{3}\right) +h_{2}\left( \sigma ^{3}\right) =\sigma ^{3}.
\label{cc}
\end{equation}%
The first set of solutions is $h_{1}\left( \sigma ^{3}\right) =\sigma
,h_{2}\left( \sigma ^{3}\right) =\sigma ^{2}$ and correspondingly $%
h_{1}\left( \sigma ^{3}\right) =\sigma ^{2}$, $h_{2}\left( \sigma
^{3}\right) =\sigma $. This means that two sets of rotation operators

\begin{eqnarray}
&&1.\quad X_{\kappa }V_{\sigma },X_{\lambda }V_{\sigma ^{2}},X_{\sigma
}V_{\sigma ^{3}},  \label{cd} \\
&&2.\quad X_{\kappa ^{\prime }}V_{\sigma },X_{\lambda ^{\prime }}V_{\sigma
^{2}},X_{\sigma ^{2}}V_{\sigma ^{3}},
\end{eqnarray}%
where $\kappa ,\kappa ^{\prime },\lambda ,\lambda ^{\prime }\in $ $GF\left(
2^{2}\right) $\ lead to the same Wigner function. Note, that the constants $%
\kappa ,\kappa ^{\prime },\lambda ,\lambda ^{\prime }$ are
arbitrary elements of $GF\left( 2^{2}\right) $ because there are
no restrictions imposed neither on $h\left( \sigma \right) $ nor
$h\left( \sigma ^{2}\right) $.

The second set of solutions of (\ref{cc}) is $h_{1}\left( \sigma ^{3}\right)
=\sigma ^{3},h_{2}\left( \sigma ^{3}\right) =0$ and, correspondingly, $%
h_{1}\left( \sigma ^{3}\right) =0,h_{2}\left( \sigma ^{3}\right) =\sigma
^{3} $, so that the rotation operators

\begin{eqnarray}
&&3.\quad X_{\kappa ^{\prime \prime }}V_{\sigma },X_{\lambda ^{\prime \prime
}}V_{\sigma ^{2}},X_{\sigma ^{3}}V_{\sigma ^{3}},  \label{ce} \\
&&4.\quad X_{\kappa ^{\prime \prime \prime }}V_{\sigma },X_{\lambda ^{\prime
\prime \prime }}V_{\sigma ^{2}},V_{\sigma ^{3}},
\end{eqnarray}%
with $\kappa ^{\prime \prime },\kappa ^{\prime \prime \prime },\lambda
^{\prime \prime },\lambda ^{\prime \prime \prime }\in $ $GF\left(
2^{2}\right) $ produce the same Wigner functions.

Finally, there are only two different Wigner functions to represent the
state $\left\vert \psi \right\rangle =\left( \left\vert 0\right\rangle
+\left\vert \sigma ^{3}\right\rangle \right) /\sqrt{2}$ (compare with \cite%
{Paz2}). It is worth noting that this state is labeled by the elements of
the field $GF\left( 2^{2}\right) $ and thus has no direct relation to the
physical state, until the basis for the field representation is fixed (see
Sec.6).

Obviously, for larger fields the variety of Wigner functions for a
given state rapidly grows with the dimension of the field even for
highly symmetrical states. For instance, the state $\left\vert
\psi \right\rangle =\left( \left\vert 0\right\rangle +\left\vert
\sigma ^{7}\right\rangle \right) /\sqrt{2}$ labelled with elements
of $GF\left( 2^{3}\right) $ can be represented in $8$ different
ways.

For fields of odd characteristic, in the whole group $\{V_{\mu }X_{\nu },\
\mu ,\nu \in GF(p^{n})\}$ a subgroup containing only rotation operators $%
\{V_{\mu },\ \mu \in GF(p^{n})\}$ can be separated, which allows us to
construct the phase space as it is outlined in previous sections, so that
the Wigner function is uniquely defined for a given state. Nevertheless, the
whole group can be used for phase-space construction as well, which would
lead to the non-uniqueness in the definition of the Wigner function, very
similar to (\ref{W_p2}).

For an arbitrary state, we can easily calculate a total number of possible
Wigner functions which represent this state in the discrete phase space.
According to the present construction we fix the phase of the state
corresponding to the horizontal line (\ref{Z_psc}). Besides, we fix the
property (\ref{ZFX}) of the Fourier transform operator (\ref{03_}), i.e. the
Fourier transformation of $Z_{\alpha }$ operators generate $X_{\alpha }$
operators without any phase factor (which in principle is not necessary if
the property $F^{4}=I$ for $d=p^{n}$, where $p\neq 2$ and $F^{2}=I$ for $%
d=2^{n}$ is not required). Now, we can generate all the possible Wigner
functions choosing different ($d-1$) rotation operators $V_{\mu }X_{\nu }$
(both for fields of odd and even characteristics), which gives $d^{d-1}$
different structures (which is directly related to different quantum nets
introduced in \cite{Wootters04}). Nevertheless, the symmetry of the state
can essentially reduce the number of different Wigner functions.

\section{Reconstruction procedure}

It is well known that the Wigner function can be reconstructed using
projective measurements, associated with a summation over the lines \cite%
{Wootters, Bjork}. In this section we explicitly relate the elements of the
density matrix with the corresponding tomogram, the averages of the form $%
\langle \psi _{\nu }^{\mu }|\rho |\psi _{\nu }^{\mu }\rangle =\omega (\mu
,\nu )$.

Let us start with the relation of the tomogram $\omega (\mu ,\nu )$ with the
Wigner function:%
\begin{equation}
\omega (\mu ,\nu )=\frac{1}{d}\sum_{\alpha ,\beta \in GF\left(
d\right) }W(\alpha ,\beta )\delta _{\beta ,\mu \alpha +\nu
}=\frac{1}{d}\sum_{\alpha \in GF\left( d\right) }W(\alpha ,\mu
\alpha +\nu ),
\end{equation}%
and consequently with the components of the density matrix in the basis of
the displacement operators (\ref{a}), which is obtained by taking into
account the relation (\ref{W_f}):%
\begin{equation*}
\omega (\mu ,\nu )=\sum_{\kappa \in GF\left( d\right) }\rho _{\kappa ,\mu
\kappa }\chi (\kappa \nu ).
\end{equation*}%
The above equation can be immediately inverted using (\ref{chi_or}):
\begin{equation*}
\rho _{\kappa ,\mu \kappa }=\frac{1}{d}\sum_{\nu \in GF\left( d\right)
}\omega (\mu ,\nu )\chi (-\kappa \nu ),
\end{equation*}%
or, changing the indices
\begin{equation*}
\rho _{\kappa ,\lambda }=\frac{1}{d}\sum_{\nu \in GF\left( d\right) }\omega
(\kappa ^{-1}\lambda ,\nu )\chi (-\kappa \nu ),
\end{equation*}%
where $\kappa \neq 0$.

To reconstruct the matrix element $\rho _{0,\lambda }$ we have to use the
results of measurements in the conjugate basis $|\widetilde{\kappa }\rangle $%
:
\begin{equation*}
\omega \left( \kappa \right) =\langle \widetilde{\kappa }|\rho |\widetilde{%
\kappa }\rangle =\frac{1}{d}\sum_{\alpha ,\beta \in GF\left( d\right)
}W(\alpha ,\beta )\delta _{\alpha ,\kappa }=\sum_{\lambda \in GF\left(
d\right) }\rho _{0,\lambda }\chi (-\kappa \lambda ),
\end{equation*}%
which leads to
\begin{equation*}
\rho _{0,\lambda }=\frac{1}{d}\sum_{\kappa \in GF\left( d\right) }\omega
(\kappa )\chi (\kappa \lambda ).
\end{equation*}

\section{Ordering the points on the axes}

In order to plot the Wigner function we have to choose an arrangement of the
field elements and take it into account to fix the order of the elements on
the axes in the finite plane $GF\left( d\right) \times GF\left( d\right) $.
When the dimension of the system is a prime number there is a natural
ordering of the elements, i.e. for any $a,b\in \mathcal{Z}_{p},a\neq b$
there is a relation either $a<b$ or $a>b$, so that the points on the axes
can be enumerated. For instance, in the case $\mathcal{Z}_{5}$ the elements
are arranged as $\left\{ 0,1,2,3,4\right\} $.

For field extensions there is no natural ordering of elements, however,
there are several possibilities to arrange elements of the field. The most
common choice is an ordering according to powers (which are natural numbers
and thus can be naturally ordered) of some primitive element. Nevertheless,
the primitive element, in general, is not unique. Actually, for the $d$%
-dimensional system there are $\phi \left( d-1\right) $ primitive elements,
where $\phi \left( r\right) $ is the Euler's function, which indicates the
number of integers with $1\leq n\leq r$ which are relatively prime to $r$
\cite{FF}.

As an another possibility we can enumerate the points on the axes according
to the following procedure: first, we fix that the origin corresponds to the
point $\left( 0,0\right) $; second, we choose some basis $\left\{ \sigma
_{1},\sigma _{2},\ldots ,\sigma _{n}\right\} $ in the field $GF(p^{n})$ and
expand elements in this basis: $\alpha =\sum_{j=1}^{n}a_{j}\sigma _{j}$, $%
a_{j}\in \mathcal{Z}_{p}.$ Now, we arrange the expansion coefficients in a
sequence to form a number on the base $p$ (binary, ternary system, etc.)
starting with the coefficient $a_{n}$ and taking the leftmost place, it will
be followed by $a_{n-1}$ and so on; obviously, $a_{1}$ takes the rightmost
place. The full number is $\left( a_{n}a_{n-1}\ldots a_{1}\right) _{p}$ and
can be transformed into some integer in the decimal basis using the standard
procedure: $\left( a_{1}\times p^{0}\right) +\left( a_{2}\times p^{1}\right)
+\ldots +\left( a_{n}\times p^{n-1}\right) $. This procedure is obviously
not unique due to the existence of different bases in the field. As an
example, let us take $GF\left( 2^{3}\right) $, the choice of the irreducible
polynomial as $x^{3}+x+1=0$, and use the normal self-dual basis $\left\{
\sigma ^{3},\sigma ^{6},\sigma ^{5}\right\} $, in this basis the field
elements are%
\begin{equation*}
\begin{array}{cc}
\theta =\theta ^{6}+\theta ^{5}, & \theta ^{2}=\theta ^{3}+\theta ^{5}, \\
\theta ^{3}=\theta ^{3}, & \theta ^{4}=\theta ^{3}+\theta ^{6}, \\
\theta ^{5}=\theta ^{5}, & \theta ^{6}=\theta ^{6}, \\
\theta ^{7}=\theta ^{3}+\theta ^{6}+\theta ^{5}, & 0.%
\end{array}%
\end{equation*}%
Now, we can associate each element with a number in the binary system%
\begin{equation*}
\begin{array}{cc}
\theta \rightarrow \left( 110\right) _{2}\rightarrow \left( 6\right) _{10},
& \theta ^{2}\rightarrow \left( 101\right) _{2}\rightarrow \left( 5\right)
_{10}, \\
\theta ^{3}\rightarrow \left( 001\right) _{2}\rightarrow \left( 1\right)
_{10}, & \theta ^{4}\rightarrow \left( 011\right) _{2}\rightarrow \left(
3\right) _{10}, \\
\theta ^{5}\rightarrow \left( 100\right) _{2}\rightarrow \left( 4\right)
_{10}, & \theta ^{6}\rightarrow \left( 010\right) _{2}\rightarrow \left(
2\right) _{10}, \\
\theta ^{7}\rightarrow \left( 111\right) _{2}\rightarrow \left( 7\right)
_{10}, & 0\rightarrow \left( 000\right) _{2}\rightarrow \left( 0\right)
_{10}.%
\end{array}%
\end{equation*}%
So, we arrange the points on the axes using the above ordering according to $%
\left\{ 0,\theta ^{3},\theta ^{6},\theta ^{4},\theta ^{5},\theta ^{2},\theta
,\theta ^{7}\right\} $.

Another possibility to arrange the elements of the field is in accordance
with the value of the trace of each element (which is a natural number) \cite%
{Klimov05b} and inside the set of the elements with the same trace we can
use any of above mentioned ordering, say powers of a primitive element.

\section{From abstract states to physical states}

In applications we have to establish a relation between abstract
states labelled with elements of the field and states of a given
physical system. Such interrelation strongly depends on the
character of a system, for instance, if the system is actually a
single "particle" with $p^{n}$ energy levels or it consists in $n$
"particles" (degrees of freedom) with $p$ energy levels. In the
last case the mapping $\mathcal{H}_{d}\Leftrightarrow
\mathcal{H}_{p}\otimes \mathcal{H}_{p}..\otimes \mathcal{H}_{p}$
from the abstract Hilbert space to $n$-particle vector space can
be achieved by expanding an element of the field in a convenient
basis $\left\{ \sigma _{1},\ldots ,\sigma _{n}\right\} $: $\alpha
=a_{1}\sigma _{1}+\ldots +a_{n}\sigma _{n}$, $a_{j}\in
\mathcal{Z}_{p}$, so that
\begin{equation*}
\left\vert \alpha \right\rangle \rightarrow \left\vert a_{1}\right\rangle
_{1}\otimes \ldots \otimes \left\vert a_{n}\right\rangle _{n}\equiv
\left\vert a_{1},\ldots ,a_{n}\right\rangle ,
\end{equation*}%
and the coefficients $a_{j}$ play the role of quantum numbers of each
particle. For instance, in the case $GF(2^{2})$ the state $\left( \left\vert
0\right\rangle +\left\vert \sigma ^{3}\right\rangle \right) /\sqrt{2}$
corresponds to the physical state $\left( \left\vert 00\right\rangle
+\left\vert 10\right\rangle \right) /\sqrt{2}$ in the polynomial basis ($%
1,\sigma $), whereas in the self-dual basis ($\sigma $,$\sigma ^{2}$) it is
associated with $\left( \left\vert 00\right\rangle +\left\vert
11\right\rangle \right) /\sqrt{2}$. Observe, that while one state is
factorizable, the other one is entangled.

This implies that all the operators $Z_{\beta }$ are factorized into a
product of single particle $Z$ operators, $Z_{\beta }=Z^{b_{1}}\otimes
\ldots \otimes Z^{b_{n}}$, where $\beta =b_{1}^{\prime }\sigma _{1}^{\prime
}+\ldots +b_{n}^{\prime }\sigma _{n}^{\prime }$ and $\left\{ \sigma
_{1}^{\prime },\ldots ,\sigma _{n}^{\prime }\right\} $ is the basis which is
dual to $\left\{ \sigma _{1},\ldots ,\sigma _{n}\right\} $, see Appendix A,
and $b_{i}^{\prime }\in \mathcal{Z}_{p}$. To have a better understanding of
this aspect let us recall the definition of the operator $Z_{\alpha }$
\begin{equation*}
Z_{\alpha }=\sum_{\beta \in GF\left( d\right) }\chi \left( \alpha \beta
\right) \left\vert \beta \right\rangle \left\langle \beta \right\vert ,
\end{equation*}%
and choose a basis $\left\{ \sigma _{1},\ldots ,\sigma _{n}\right\} $ to
expand $\beta $. Then, taking $\alpha =\sigma _{i}^{\prime }$ as an element
of the dual basis, we obtain%
\begin{equation*}
Z_{\sigma _{i}^{\prime }}=\sum_{\beta \in GF\left( d\right) }\chi \left(
\sigma _{i}^{\prime }\beta \right) \left\vert \beta \right\rangle
\left\langle \beta \right\vert =\Pi _{j=1}^{n}\sum_{b_{j}=0}^{p-1}\exp
\left( \frac{2\pi i}{p}b_{j}\mathrm{tr}\left( \sigma _{i}^{\prime }\sigma
_{j}\right) \right) \left\vert b_{j}\right\rangle \left\langle
b_{j}\right\vert .
\end{equation*}%
Now, if a) $i\neq j$ \ the duality means $\mathrm{tr}\left( \sigma
_{i}^{\prime }\sigma _{j}\right) =0$, and thus
\begin{equation*}
\sum_{b_{j}=0}^{p-1}\exp \left( \frac{2\pi i}{p}b_{j}\mathrm{tr}\left(
\sigma _{i}^{\prime }\sigma _{j}\right) \right) \left\vert
b_{j}\right\rangle \left\langle b_{j}\right\vert
=\sum_{b_{j}=0}^{p-1}\left\vert b_{j}\right\rangle \left\langle
b_{j}\right\vert =I_{j},
\end{equation*}%
where the index $j$ means the $j$-th particle;

b) $i=j$ we have $\mathrm{tr}\left( \sigma _{i}^{\prime }\sigma _{i}\right)
=1$ and then,%
\begin{equation*}
\sum_{b_{i}=0}^{p-1}\exp \left( \frac{2\pi i}{p}b_{i}\mathrm{tr}\left(
\sigma _{i}^{\prime }\sigma _{i}\right) \right) \left\vert
b_{i}\right\rangle \left\langle b_{i}\right\vert =\sum_{b_{i}=0}^{p-1}\exp
\left( \frac{2\pi i}{p}b_{i}\right) \left\vert b_{i}\right\rangle
\left\langle b_{i}\right\vert =Z_{i}.
\end{equation*}%
Finally, we obtain the factorization
\begin{equation*}
Z_{\sigma _{i}^{\prime }}=I_{1}\otimes \ldots \otimes I_{i-1}\otimes
Z_{i}\otimes I_{i+1}\otimes \ldots I_{n},
\end{equation*}%
i.e. in the $i$-th place is located $Z$ operator and all the others are
unities.

In the case of a single "particle" the states of a physical
systems can be labelled by the elements of the field arranged in
some order ($0,\alpha _{1},...,\alpha _{p^{n-1}}$) (see previous
Section). Then, the free
Hamiltonian of the systems takes on the form%
\begin{equation*}
H=E_{0}\left\vert 0\right\rangle \left\langle 0\right\vert +E_{1}\left\vert
\alpha _{1}\right\rangle \left\langle \alpha _{1}\right\vert +\ldots
+E_{p^{n}}|\alpha _{p^{n-1}}\rangle \langle \alpha _{p^{n-1}}|,
\end{equation*}%
where the energies are arranged in the non-decreasing order: $E_{0}\leq
E_{1}\leq \ldots \leq E_{p^{n}}$.

\section{Conclusions}

In this paper we studied an explicit form of the kernel operator
(the phase point operator \cite{Wootters,Wootters04}) which maps
states of a quantum system of dimension $d=p^{n}$ into a Wigner
function in a discrete phase space. The crucial point in the phase
space construction play the rotation and displacement operators
labelled with elements of $GF(d)$. These operators are resulted
explicitly related after imposing the condition (\ref{W_rho_1})
and allow us to establish a clear correspondence between states in
the Hilbert space of the system and lines in the discrete phase
space. The structure of the rotation operators is quite different
for fields of odd and even characteristic. While for the fields of
odd characteristic the rotation operators form a $p^{n}$
dimensional Abelian group, the corresponding group in the case of
$GF(2^{n})$ is of order $2^{2n}$ and includes both "rotation" and
"vertical" displacement operators $X_{\mu },\mu \in GF(2^{n})$. So
that although, for a particular phase-space construction the group
property is not really unnecessary, different choices of sets of
rotation operators, lead to different Wigner functions, which is
directly connected to the
freedom in the election of quantum nets in the Wootters' construction \cite%
{Wootters04}. Such freedom obviously exists also in the case of odd
characteristics, which nevertheless can be avoided fixing the rotation group
in a natural way (\ref{vexplicit}).

\textbf{Acknowledgements}

This work is partially supported by the Grant 45704 of Consejo
Nacional de Ciencia y Tecnologia (CONACyT).

\section{Appendix A: Finite fields}

A set $\mathcal{L}$ is a commutative \textit{ring} if two binary operations:
addition and multiplication (both commutative and associative) are defined.

A \textit{field} $F$ is a commutative ring with division, i.e. for any $a\in
F$ there exists $a^{-1}\in F$ so that $a^{-1}a=aa^{-1}=I$ (excluding the
zero element). Elements of a field form groups with respect to addition $F$
and multiplication $F^{\ast }=F-\left\{ 0\right\} $.

The characteristic of a \textit{finite field} is the smallest
integer $p$, so that $p\cdot
1=\underbrace{1+1+..+1}_{\mbox{\scriptsize $p$ times}}=0$ and is
always a prime number. Any finite field contains a prime subfield
$\mathcal{Z}_{p}$ and has $p^{n}$ elements, where $n$ is a natural
number. Moreover, the finite field containing $p^{n}$ elements is
unique and is usually called
Galois field, $GF(p^{n})$. $GF(p^{n})$ is an extension of degree $n$ of $%
\mathcal{Z}_{p}$, i.e. elements of $GF(p^{n})$ can be obtained with $%
\mathcal{Z}_{p}$ and all the roots of an \textit{irreducible polynomial}
(That is, one which cannot be factored in $\mathcal{Z}_{p}$) with
coefficients inside $\mathcal{Z}_{p}$.

The multiplicative group of $GF(p^{n}):GF(p^{n})^{\ast }=GF(p^{n})-\left\{
0\right\} $ is cyclic $\theta ^{p^{n}}=\theta $, $\theta \in GF(p^{n})$. The
generators of this group are called \textit{primitive elements} of the field.

A primitive element of $GF(p^{n})$ is a root of an irreducible polynomial of
degree $n$ over $\mathcal{Z}_{p}$. This polynomial is called a \textit{%
minimal polynomial}.

The trace operation
\begin{equation*}
\mathrm{tr}(\alpha )=\alpha +\alpha ^{2}+...+\alpha ^{p^{n-1}},
\end{equation*}%
maps any field element into an element of the prime field, $\mathrm{tr}:%
\underset{\alpha }{GF(p^{n})}\rightarrow \underset{\mathrm{tr}\left( \alpha
\right) }{\mathcal{Z}_{p}}$, and satisfies the property%
\begin{equation}
\mathrm{tr}(\alpha _{1}+\alpha _{2})=\mathrm{tr}(\alpha _{1})+\mathrm{tr}%
(\alpha _{2}).  \label{tracesum}
\end{equation}%
The additive characters are defined as
\begin{equation*}
\chi (\alpha )=\exp \left[ \frac{2\pi i}{p}\mathrm{tr}\left( \alpha \right) %
\right] ,
\end{equation*}%
and possess two important properties:
\begin{equation*}
\chi (\alpha _{1}+\alpha _{2})=\chi (\alpha _{1})\chi (\alpha _{2})
\end{equation*}%
and
\begin{equation*}
\sum_{\alpha \in GF\left( p^{n}\right) }\chi (\alpha )=0.
\end{equation*}

Any finite field $GF(p^{n})$ can be considered as an $n$-dimensional linear
vector space and there is a basis $\{\sigma _{j},j=1,..,n\}$ in this vector
space, so that any $\alpha \in GF(p^{n})$, $\alpha
=\sum_{j=1}^{n}a_{j}\sigma _{j}$ and $a_{j}\in \mathcal{Z}_{p}$. There are
several bases, one of them is the \textit{polynomial basis} $\{1,\theta
,\theta ^{2},...,\theta ^{n-1}\}$, where $\theta $ is a primitive element of
$GF(p^{n})$, another one is the \textit{normal basis }$\{\theta ,\theta
^{p},...,\theta ^{p^{n-1}}\}$, so one can choose whatever according to the
specific problem.

Two bases $\left\{ \alpha _{1},\ldots ,\alpha _{n}\right\} $ and $\left\{
\beta _{1},\ldots ,\beta _{n}\right\} $ in the same field are dual if $%
\mathrm{tr}\left( \alpha _{i}\beta _{j}\right) =\delta _{ij}.$A basis which
is dual to itself is called \textit{self-dual basis}, $\mathrm{tr}\left(
\alpha _{i}\alpha _{j}\right) =\delta _{ij}$.

Example: $GF\left( 2^{2}\right) $, the primitive polynomial is $x^{2}+x+1=0$%
, it has the roots $\left\{ \theta ,\theta ^{2}\right\} $. The polynomial
basis is $\left\{ 1,\theta \right\} $, whose dual basis is $\left\{ \theta
^{2},1\right\} $:%
\begin{equation*}
\begin{array}{cc}
\mathrm{tr}\left( 1\theta ^{2}\right) =1, & \mathrm{tr}\left( 11\right) =0,
\\
\mathrm{tr}\left( \theta \theta ^{2}\right) =0, & \mathrm{tr}\left( \theta
1\right) =1.%
\end{array}%
\end{equation*}%
The normal basis $\left\{ \theta ,\theta ^{2}\right\} $ is self-dual:%
\begin{equation*}
\begin{array}{cc}
\mathrm{tr}\left( \theta \theta \right) =1, & \mathrm{tr}\left( \theta
\theta ^{2}\right) =0, \\
\mathrm{tr}\left( \theta ^{2}\theta \right) =0, & \mathrm{tr}\left( \theta
^{2}\theta ^{2}\right) =1.%
\end{array}%
\end{equation*}

\section{Appendix B: Solution of the equation $c_{\protect\kappa +\protect%
\alpha ,\protect\mu }c_{\protect\kappa ,\protect\mu }^{\ast }=c_{\protect%
\alpha ,\protect\mu }\protect\chi \left( \protect\mu \protect\alpha \protect%
\kappa \right) $ in the $d=2^{n}$ case}

To solve Eq. (\ref{ck_2}) we first fix a basis in the field $GF(2^{n})$: $%
\{\sigma _{j},\ j=1,..,n\},$ so that any element of the field can be
represented as a linear combination
\begin{equation}
\alpha =\sum_{j=1}^{n}a_{j}\sigma _{j},\quad a_{j}\in \mathcal{Z}_{2}.
\label{expan}
\end{equation}%
Then, we solve the equation (\ref{c_p2}) $c_{\kappa ,\mu }^{2}=\chi \left(
\kappa ^{2}\mu \right) $ for the $n$ basis elements $c_{\kappa ,\mu }$, $%
\kappa =\sigma _{1},...,\sigma _{n}$ assigning in an arbitrary way
the signs $\pm 1$ to the square root $\sqrt{\chi \left( \kappa
^{2}\mu \right) }$. This means that there exist $2^{n}$ different
sets of $\left\{ c_{\kappa ,\mu }\right\} $, and thus $2^{n}$
different operators $V_{\mu }$ (for a fixed value of $\mu $),
which can be precisely parameterized as in Eq.(\ref {VV}). Once
the signs of $c_{\kappa ,\mu }$, $\kappa =\sigma _{1},...,\sigma
_{n}$, are fixed, the rest of the $2^{n}-n$ coefficients
$c_{\kappa ,\mu }$ can be found using the expansion (\ref{expan})
and the relation (\ref{ck_2}) in the form $c_{\kappa +\alpha ,\mu
}=c_{\kappa ,\mu }c_{\alpha ,\mu }\chi
\left( -\mu \alpha \kappa \right) $ leading to the following result:%
\begin{eqnarray*}
c_{\alpha ,\mu } &=&c_{a_{1}\sigma _{1}+..+a_{n}\sigma _{n},\mu
}=c_{a_{1}\sigma _{1},\mu }c_{a_{2}\sigma _{2}+..+a_{n}\sigma _{n},\mu }\chi
\left( a_{1}\sigma _{1}\left( a_{2}\sigma _{2}+..+a_{n}\sigma _{n}\right)
\mu \right) = \\
&=&...=\chi \left( \mu \sum_{k=1}^{n-1}a_{k}\sigma
_{k}\sum_{j=k+1}^{n}a_{j}\sigma _{j}\right) \Pi _{l=1}^{n}c_{a_{l}\sigma
_{l},\mu },
\end{eqnarray*}%
and $c_{0,\mu }=1$.

To illustrate how this procedure works let us apply it to the case of $%
GF(2^{2})$. We choose the normal, self-dual, basis \{$\theta ,\theta ^{2}$\}
(see Appendix A) in $GF(2^{2})$, where $\theta $ is a root of the primitive
polynomial $x^{2}+x+1=0$, so that $\theta ^{3}=\theta +\theta ^{2}$. The
solution of Eq. (\ref{c_p2}) for, say $\mu =\theta ^{3}=1$, is (below we
will omit the index $\mu $ in the coefficients $c_{\alpha ,\mu }$)%
\begin{equation*}
c_{0}=1,\;c_{\theta }=\pm i,\;c_{\theta ^{2}}=\pm i.
\end{equation*}%
Then, the last coefficient is given by%
\begin{equation*}
c_{\theta ^{3}}=c_{\theta +\theta ^{2}}=c_{\theta }c_{\theta ^{2}}\chi
\left( -\theta \right) =\left( \pm i\right) \left( \pm i\right) \left(
-1\right) ,
\end{equation*}%
and one can see that there exist $4$ different possible operators $V_{\theta
^{3}}$. A similar calculus can be made for the operators $V_{\theta }$ and $%
V_{\theta ^{2}}$.

It is convenient to fix positive signs of the coefficients $c_{\kappa ,\mu }$
corresponding to the elements of the field basis, i.e. $c_{a_{l}\sigma
_{l},\mu }=\sqrt{\chi \left( a_{l}^{2}\sigma _{l}^{2}\mu \right) }$, $%
l=1,..,n$, and form the "first" set of the rotation operators,
$V_{\mu }$ with\ these coefficients . Then, all the other sets of
$V_{\mu ,\nu }$ can be obtained according to (\ref{VV}).

Once we have the coefficients $c_{\alpha ,\mu }$ one can easily obtain the
corresponding phase factors for the displacement operator (\ref{phi_gen}).
For instance, fixing the "first" set of rotation operators in the above
example as

\begin{equation}
V_{\theta }=diag(1,1,i,-i),\quad V_{\theta ^{2}}=diag(1,i,1,-i),\quad
V_{\theta ^{3}}=diag(1,i,i,-1),  \label{++V}
\end{equation}%
we obtain the following phase factors appearing in the displacement operator:

\begin{equation*}
\begin{tabular}{lll}
$\phi (\theta ,\theta )=i,$ & $\phi (\theta ,\theta ^{2})=1,$ & $\phi
(\theta ,\theta ^{3})=i,$ \\
$\phi (\theta ^{2},\theta )=1,$ & $\phi (\theta ^{2},\theta ^{2})=i,$ & $%
\phi (\theta ^{2},\theta ^{3})=i,$ \\
$\phi (\theta ^{3},\theta )=-i,$ & $\phi (\theta ^{3},\theta ^{2})=-i,$ & $%
\phi (\theta ^{3},\theta ^{3})=-1.$%
\end{tabular}%
\end{equation*}%
Another set of rotation operators can be obtained, for instance, keeping $%
V_{\theta }$ and $V_{\theta ^{2}}$ as in the above and multiplying the
operator $V_{\theta ^{3}}$ in (\ref{++V}) by $X_{\theta }$: $\ $
\begin{equation*}
V_{\theta }=diag(1,1,i,-i),\quad V_{\theta ^{2}}=diag(1,i,1,-i),\quad
V_{\theta ^{3}}=diag(1,-i,i,1),
\end{equation*}%
which leads to some changes in the phases of the displacement operator:

\begin{equation*}
\begin{tabular}{lll}
$\phi (\theta ,\theta )=-i,$ & $\phi (\theta ,\theta ^{2})=1,$ & $\phi
(\theta ,\theta ^{3})=i,$ \\
$\phi (\theta ^{2},\theta )=1,$ & $\phi (\theta ^{2},\theta ^{2})=i,$ & $%
\phi (\theta ^{2},\theta ^{3})=i,$ \\
$\phi (\theta ^{3},\theta )=-i,$ & $\phi (\theta ^{3},\theta ^{2})=-i,$ & $%
\phi (\theta ^{3},\theta ^{3})=1.$%
\end{tabular}%
\end{equation*}

\section{Appendix C: Determination of $f\left( \protect\mu ,\protect\mu %
^{\prime }\right) $}

In practice, the function $f\left( \mu ,\mu ^{\prime }\right) $ is
determined by solving the following equation

\begin{equation}
c_{\alpha ,\mu }c_{\alpha ,\mu ^{\prime }}=\chi \left( \alpha f\left( \mu
,\mu ^{\prime }\right) \right) c_{\alpha ,\mu +\mu ^{\prime }}
\label{cc_chi}
\end{equation}%
for all values of $\alpha $ belonging to the basis of the field. The rest of
the elements of the field do not provide additional information due to (\ref%
{ck_2}).

As an example we consider the case of $GF(2^{2})$. Following the general
procedure, we choose the self-dual basis ($\theta ,\theta ^{2}$) in the
field and fix the rotation operators as in (\ref{++V}). The equation (\ref%
{cc_chi}) for $f\left( \theta ,\theta \right) $ has the form

\begin{equation*}
c_{\alpha ,\theta }c_{\alpha ,\theta }=\chi \left( \alpha f\left( \theta
,\theta \right) \right) c_{\alpha ,0},
\end{equation*}%
and for different values of the parameter $\alpha $ we get

\begin{eqnarray*}
\alpha &=&\theta ,\qquad 1\cdot 1=\chi \left( \theta f\left( \theta ,\theta
\right) \right) 1, \\
\alpha &=&\theta ^{2},\qquad i\cdot i=\chi \left( \theta ^{2}f\left( \theta
,\theta \right) \right) 1,
\end{eqnarray*}%
leading to the only possible solution $f\left( \theta ,\theta \right)
=\theta ^{2}$.

Similarly we obtain for the rest of the $f\left( \mu ,\mu ^{\prime }\right) $%
:

for $f\left( \theta ,\theta ^{2}\right) $

\begin{eqnarray*}
\alpha =\theta ,\qquad 1\cdot i=\chi \left( \theta f\left( \theta ,\theta
^{2}\right) \right) i, \\
\alpha =\theta ^{2},\qquad i\cdot 1=\chi \left( \theta ^{2}f\left( \theta
,\theta ^{2}\right) \right) i,
\end{eqnarray*}%
so that $f\left( \theta ,\theta ^{2}\right) =0$, and following the same idea
we can determine

\begin{equation*}
f\left( \theta ,\theta ^{3}\right) =\theta ^{2},\ f\left( \theta ^{2},\theta
^{2}\right) =\theta ,\ f\left( \theta ^{2},\theta ^{3}\right) =\theta ,\
f\left( \theta ^{3},\theta ^{3}\right) =\theta ^{3}.
\end{equation*}

\section{Appendix D: Another symplectic operators}

\subsection{$U$ operator}

In order to introduce another operator \cite{Vourdas} with similar
properties to $V_{\mu }$, let us define that two lines are orthogonal if the
states corresponding to these ones are related via the Fourier transform%
\begin{equation*}
|\kappa \rangle \overset{F}{\rightarrow }|\widetilde{\kappa }\rangle .
\end{equation*}%
Then, there exists an operator $U_{\mu }|\widetilde{\kappa }\rangle =FV_{\mu
}|\kappa \rangle $ such that its geometrical application rotates the line
corresponding to the state $|\kappa \rangle $ (conjugate to $|\widetilde{%
\kappa }\rangle $) and then, transforms the rotated line into an orthogonal
one. This operator is obtained from $V_{\mu }$ as
\begin{equation}
U_{\mu }=FV_{\mu }F^{\dagger }=\sum\limits_{\kappa \in GF\left(
d\right) }c_{-\kappa ,\mu }|\kappa \rangle \langle \kappa |,
\label{u}
\end{equation}%
so that, $U_{\mu }Z_{\alpha }U_{\mu }^{\dagger }=Z_{\alpha }$ for all $%
\alpha ,\mu $. The action of $U_{\mu }$ on the operator $X_{\beta
}$ can be obtained using the operational relation in (\ref{u}) and
the relation (\ref{ZFX})
\begin{equation}
U_{\mu }X_{\beta }U_{\mu }^{\dagger }=F(V_{\mu }Z_{\beta }V_{\mu }^{\dagger
})^{\dagger }F^{\dagger }=\exp (-i\varphi \left( \beta ,\mu \right) )Z_{\mu
\beta }^{\dagger }X_{\beta },  \label{ux}
\end{equation}%
which means that the $U$-transformation also represents a sort of
rotation: operators $U_{\mu ^{\prime }}$ transform eigenstates of
the set of displacement operators labelled with points of the ray
$\beta =\mu \alpha $

\begin{equation}
\left\{ I,Z_{\alpha _{1}}X_{\mu \alpha _{1}},Z_{\alpha _{2}}X_{\mu \alpha
_{2}},...\right\}
\end{equation}%
to the eigenstates of the set labelled with points of the ray
$\beta =\left( \mu +\mu ^{\prime }\right) ^{-1}\alpha $

\begin{equation}
\left\{ I,Z_{\left( 1+\mu \mu ^{\prime }\right) \alpha _{1}}X_{\mu \alpha
_{1}},Z_{\left( 1+\mu \mu ^{\prime }\right) \alpha _{2}}X_{\mu \alpha
_{2}},...\right\} .
\end{equation}%
Similarly, as we cannot reach the ray $\alpha =0$ using $V_{\mu }$
operators, one cannot reach the ray $\beta =0$ using $U_{\mu ^{\prime }}$
operators. Having both operators $U_{\mu ^{\prime }}$ and $V_{\mu }$ we can
transform any ray into any other.

Let us find the Wigner function of a state transformed by $V_{\mu }$ and $%
U_{\mu }$ operators. For fields of odd characteristic, $p\neq 2$, one can
use the explicit form (\ref{vexplicit}) of $V_{\mu }$ and after some algebra
we obtain%
\begin{equation}
W_{\tilde{\rho}}\left( \alpha ,\beta \right) =W_{\rho }\left( \alpha ,\beta
-\mu \alpha \right) ,\quad \tilde{\rho}=V_{\mu }\rho V_{\mu }^{\dagger }.
\label{WV}
\end{equation}%
In the same manner we evaluate the Wigner function of\ a state transformed
by $U_{\mu }$ operator, $p\neq 2$:%
\begin{equation}
W_{\tilde{\rho}}\left( \alpha ,\beta \right) =W_{\rho }\left( \alpha -\mu
\beta ,\beta \right) ,\quad \tilde{\rho}=U_{\mu }\rho U_{\mu }^{\dagger }.
\label{WU}
\end{equation}%
In other words, the transformation of a state by $V_{\mu }$ and/or $U_{\mu }$
operators leads to a covariant transformation of the Wigner function for the
fields of odd characteristics.

\subsection{$S$ operator}

The last operator which transforms the operators labelled with
points of one line into the operators labelled with the points of
some other line is the
so-called\ squeezing operator, which in the basis of eigenstates of $%
Z_{\alpha }$ operators has the following form

\begin{equation}
S_{\xi }=\sum\limits_{\kappa \in GF\left( p^{n}\right) }\left\vert
\kappa \right\rangle \left\langle \xi \kappa \right\vert .
\label{s3}
\end{equation}%
It is easy \ to see that
\begin{equation*}
W_{\tilde{\rho}}\left( \alpha ,\beta \right) =W_{\rho }\left( \xi \alpha
,\xi ^{-1}\beta \right) ,\quad \tilde{\rho}=S_{\xi }\rho S_{\xi }^{\dagger }.
\end{equation*}%
The squeezing operator naturally appears as a result of consecutive
application of $V_{\mu }$ and $U_{\nu }$:
\begin{eqnarray*}
W_{\rho ^{\prime }} &=&W_{S_{\xi }\rho S_{\xi }^{\dag }}, \\
\tilde{\rho} &=&V_{-\xi \left( \xi -1\right) \mu ^{-1}}U_{-\xi ^{-1}\mu
}V_{\left( \xi -1\right) \mu ^{-1}}U_{\mu }\rho U_{\mu }^{\dag }V_{\left(
\xi -1\right) \mu ^{-1}}^{\dag }U_{-\xi ^{-1}\mu }^{\dag }V_{-\xi \left( \xi
-1\right) \mu ^{-1}}^{\dag },
\end{eqnarray*}%
where $\mu $ is an arbitrary element of $GF\left( p^{n}\right) ^{\ast }$, $%
p\neq 2$.

\end{document}